\documentclass{aa}  
\usepackage{graphicx}
\usepackage[english]{babel}
\usepackage{amsmath}
\usepackage{txfonts}

\begin{document}

\title{
Bandwidth in bolometric interferometry}
\subtitle{}
\author{R. Charlassier\inst{1} \and E. F. Bunn\inst{2} \and J.-Ch. Hamilton\inst{1} \and J. Kaplan\inst{1} \and S. Malu\inst{3}}

\offprints{\tt rcharlas@apc.univ-paris7.fr}

   \institute{APC, Universit\'e Denis Diderot-Paris 7, CNRS/IN2P3, CEA, Observatoire de Paris; 10 rue A. Domon \& L. Duquet, Paris, France
    \and Physics Department, University of Richmond; Richmond, VA 23173, USA \and Inter-University Centre for Astronomy and Astrophysics (IUCAA); Pune - 411 007, India}

   \date{Received ; accepted }
 
  \abstract 
 {Bolometric Interferometry is a technology currently under
 development that will be first dedicated to the detection of B-mode
 polarization fluctuations in the Cosmic Microwave Background. A
 bolometric interferometer will have to take advantage of the wide
 spectral detection band of its bolometers in order to be competitive
 with imaging experiments. A crucial concern is that interferometers
 are presumed to be importantly affected by a spoiling effect known as
 bandwidth smearing.}
	{In this paper, we investigate how the bandwidth modifies the work principle of a bolometric interferometer and how it affects its sensitivity to the CMB angular power spectra.}
{We obtain analytical expressions for the broadband visibilities
measured by broadband heterodyne and bolometric interferometers.  We
investigate how the visibilities must be reconstructed in a broadband
bolometric interferometer and show that this critically depends on
hardware properties of the modulation phase shifters. If the phase shifters
produce shifts that are
constant with respect to frequency, the instrument works
exactly as a monochromatic one (the modulation matrix is not
modified), while if they vary (linearly or otherwise) 
with respect to frequency, one
has to perform a special reconstruction scheme, which allows 
the visibilities to be reconstructed in frequency sub-bands. Using an angular
power spectrum estimator accounting for the bandwidth, we finally
calculate the sensitivity of a broadband bolometric interferometer. A
numerical simulation has been performed and confirms the analytical
results.}
{We conclude (i) that broadband bolometric interferometers allow 
broadband visibilities to be reconstructed 
whatever the kind of phase shifters
used and (ii) that for dedicated B-mode bolometric interferometers,
the sensitivity loss due to bandwidth smearing is quite acceptable,
even for wideband instruments (a factor 2 loss for a typical 20\%
bandwidth experiment).}
	{}
	

\maketitle

\section*{Introduction}
The detection of primordial gravity waves through B-mode polarization
anisotropies in the Cosmic Microwave Background (CMB) is one of the
most exciting challenges of modern cosmology. The weakness of the
expected signal requires the development of highly sensitive
experiment with an exquisite control of systematic errors. Most of the
experiments or projects dedicated to this quest are based on 
well-known direct imaging technology. An appealing alternative called
bolometric interferometry has been proposed [\cite{BI}]. This
technology combines the advantages of interferometry in handling
systematic effects and those of bolometric detectors in 
sensitivity. The two teams that have taken up the challenge
[\cite{MBI}; \cite{BrainRomain}] are now joining their efforts within
the QUBIC collaboration [\cite{qubicjean}].

We have introduced in [\cite{coherentsummation}, hereafter C08] a
simple formalism describing the general design of a bolometric
interferometer working at a monochromatic frequency and shown that a
bolometric interferometer must follow a particular phase shifting
scheme we called ``coherent summation of equivalent baselines''. This
scheme has been optimized further in [\cite{hyland}]. We have
calculated in [\cite{BIsensitivity}, hereafter H08] the sensitivity of
such a bolometric interferometer and shown that this technology can be
competitive with imaging experiments and heterodyne interferometers
for the measurement of CMB B-mode. For the sake of simplicity, we did
not deal with the bandwidth question in [C08] and [H08].

We know that a dedicated B-mode bolometric interferometer will have to
use the wide spectral detection band of its bolometers in order to be
competitive with imaging experiments. In the other hand, the bandwidth
is often considered as a crucial issue in radio-interferometry; if the
raw sensitivity of radio interferometer detectors grows as the square
root of the bandwidth, there is a secondary effect, well known as
\textit{bandwidth smearing}, which can largely degrade the global
sensitivity. When the signals coming from a point source interfere
after having been collected by two broadband receivers, the resulting
fringe pattern is smeared by an envelope whose amplitude depends on
the bandwidth, consequently leading to a degradation of the signal to
noise ratio -- see for instance~[\cite{Thomson}]. We will see that these
two main characteristics remain in bolometric interferometry: the
bolometers' sensitivity also grows as the square root of the bandwidth,
and a bandwidth smearing of the observables, the visibilities,
degrades the global sensitivity of the instrument (however, because the
observation of CMB angular correlations requires a poorer spatial
resolution than the observation of point sources to which classical
radio-interferometers are mostly dedicated, this smearing will lead to
a less critical sensitivity loss). But we will also see that an
additional kind of bandwidth issue occurs, due to the fact that in
bolometric interferometry, visibilities are not measured
\textit{directly} but by solving a linear problem.

We investigate how the visibilities are smeared in heterodyne and
bolometric interferometers having wide spectral bands and large
primary beams in section~\ref{sec:broadvis}. We investigate how the
work principle of bolometric interferometry is affected by bandwidth
in section~\ref{sec:broadbolvis}.  We show in particular that the
visibilities can be reconstructed exactly as in the monochromatic case
detailed in [C08] if the modulation phase shifts are constant with
respect to frequency, while one has to perform the special
reconstruction scheme described in section~\ref{sec:virtsub} when the
modulation phase shifts vary with respect to frequency. In
section~\ref{sec:sensitivity}, we introduce an angular power spectrum
estimator accounting for the bandwidth and estimate how the bandwidth
smearing results in a degradation of the sensitivity for B-mode
experiments. A numerical simulation that confirms our analytical
results is presented in section~\ref{sec:mcsimu}.
 
\section{Visibilities measured by generic interferometers with wide spectral bands and large primary beams \label{sec:broadvis}}

\subsection{Monochromatic visibilities}

The observables measured by a monochomatic interferometer working at a
frequency $\nu_{0}$ and looking at a radiation field of spectral power
$I_{\nu}(\vec{n})$, in units of {[}$\mathrm{W.Hz^{-1}.sr^{-1}}${]},
are called the visibilities. A \textit{monochromatic visibility} is
defined for one baseline $\vec{u}_{0}$, which is the vector separation
between two horns in units of the electromagnetic wavelength of the
radiation:
\begin{equation}
V_{I}^{\nu_{0}}(\vec{u_{0}})=\int I_{\nu_{0}}(\vec{n})A_{\nu_{0}}(\vec{n})\ \mathrm{exp}(i2\pi\vec{u_{0}}\cdot\vec{n})\ d\vec{n},
\end{equation}
where $A_{\nu}(\vec{n})=B_{\nu}^{2}(\vec{n})$ is the square of the
beam of the input horns (assumed identical), conventionally normalized to
one at its maximum.  Here we can make a first important
observation for the understanding of bandwidth in interferometry: a
monochromatic visibility defined by a pair of horns separated by a
distance $\vec{d_{0}}$ and working at frequency $\nu_{0}$ is the
same observable as a monochromatic visibility defined by another
pair of horns separated by a distance
$\vec{d}=\vec{d_{0}}\frac{\nu_{0}}{\nu}$ and working at 
frequency $\nu$. The two visibilities indeed match the baseline
$\vec{u_{0}}=\vec{d_{0}}\frac{\nu_{0}}{c}$.  This property, sometimes
called \textit{the equivalence theorem} of interferometry, is actually
true only if the two pairs of horns have the same beam (meaning that
their surfaces are necessarily different) and if the observed
radiation field $S(\vec{n})$ has the same spatial variations at both
frequencies (this is of course true in the case of CMB observations).

\subsection{Broadband visibilities for a generic interferometer}

We consider now an interferometer that is sensitive to a finite spectral
band through a bandpass function $J(\nu-\nu_0)$, centered%
\footnote{The definition of the center is somewhat arbitrary.  A convenient
definition is the barycenter of $J$.%
} at frequency $\nu_{0}$. We arbitrarily define
the \textit{bandwidth} $\Delta\nu$ of the instrument as%
\footnote{This definition is very close to the FWHM for a Gaussian bandwidth.%
}: 
\begin{equation}
\Delta\nu=\frac{1}{J(0)}\int J(\nu-\nu_0)d\nu.
\end{equation}
We define a \textit{generic interferometer} to be
an instrument in which visibilities
are directly given by the outputs of the detectors (this is the case
in heterodyne interferometry, but not in bolometric interferometry).
The expression of a \textit{broadband visibility} measured by a generic
interferometer -- 
in power units, for a baseline $\vec{u_{0}}=\vec{d_{0}}\frac{\nu_{0}}{c}$ --
is then: 
\begin{equation}
V_{I}^{\Delta\nu}(\vec{u_{0}})=\iint I_{\nu}(\vec{n})A_{\nu}(\vec{n})\ \mathrm{exp}(i2\pi\vec{d_{0}}\cdot\vec{n}\frac{\nu}{c})\ J(\nu-\nu_{0})\ d\nu\ d\vec{n}.\label{eq:broadvispower}
\end{equation}

The baselines define a plane usually called the \textit{uv-plane}.
It is better to write the visibility as a convolution in the uv-plane
to understand the bandwidth effect: 
\begin{equation}
V_{I}^{\Delta\nu}(\vec{u_{0}})=\iiiint\tilde{I}_{\nu}(\vec{w})\tilde{A}_{\nu}(\vec{w'})J(\nu-\nu_{0})e^{i2\pi(\frac{\nu}{\nu_{0}}\vec{u}_{0}-\vec{w}-\vec{w'})
\cdot\vec{n}}d\vec{w}d\vec{w'}d\vec{n}d\nu,\end{equation}
where we have introduced the Fourier transform of the signal and
the one of the beam: 
\begin{equation}
I_{\nu}(\vec{n})=\int\tilde{I}_{\nu}(\vec{w})\exp(-i2\pi\vec{w}\cdot\vec{n})d\vec{w}
\end{equation}
 \begin{equation}
A_{\nu}(\vec{n})=\int\tilde{A}_{\nu}(\vec{w'})\exp(-i2\pi\vec{w'}\cdot
\vec{n})d\vec{w'}
\end{equation}
 In the flat-sky approximation, the integral over the field $\vec{n}$
gives a delta function, and the expression of the broadband visibility
is finally 
\begin{equation}
V_{I}^{\Delta\nu}(\vec{u_{0}})=J(0)\ \Delta\nu\int\tilde{I}_{\nu_{0}}(\vec{w})\tilde{\beta}(\vec{u_{0}},\vec{w})\ d\vec{w}, 
\end{equation}
 where we have defined the \textit{convolution kernel} in the uv-plane:
\begin{equation}
\tilde{\beta}(\vec{u_{0}},\vec{w})=\int\frac{\tilde{I}_{\nu}(\vec{w})}{\tilde{I}_{\nu_{0}}(\vec{w})}\tilde{A}_{\nu}(\vec{u_{0}}\frac{\nu}{\nu_{0}}-\vec{w})\ J_{N}(\nu-\nu_{0})\ d\nu,
\end{equation}
in which we have introduced the normalized bandpass function
$J_{N}(\nu-\nu_0)=J(\nu-\nu_0)/\Delta\nu$.  
The convolution kernel (which depends
on $\vec{u_{0}}$) contains the entire effect of the bandwidth
smearing. For a monochromatic visibility, the convolution kernel in
the uv-plane is just the Fourier transform of the beam
$\tilde{A}(\vec{w})$.

\begin{figure*}[!ht]
\resizebox{\hsize}{!}{\centering{ \includegraphics{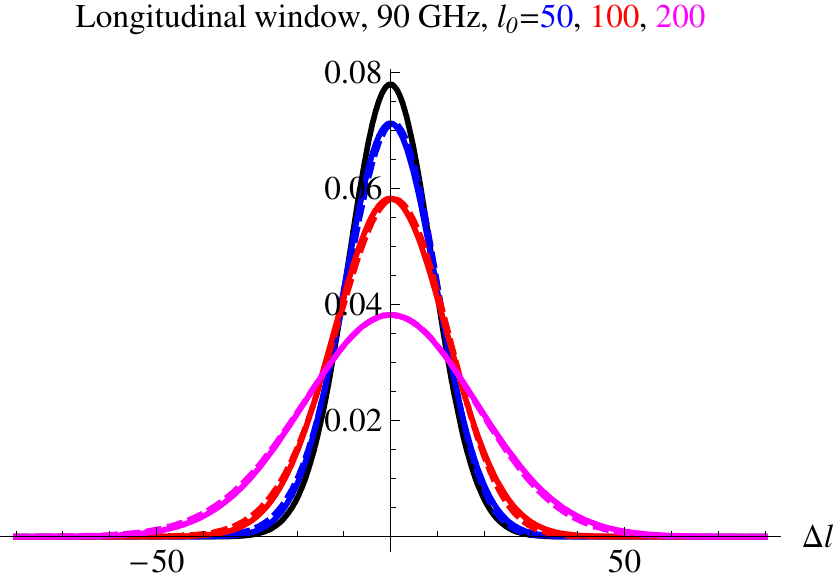}\includegraphics{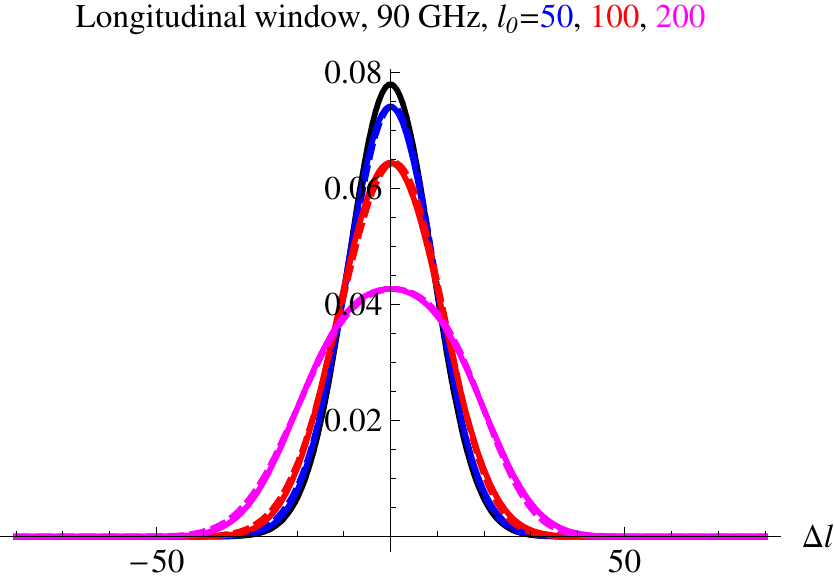}}} 

\caption{Left: longitudinal profile of the convolution kernel for baselines
corresponding to 
$l_0=50, 100, 200$, for an instrument with a 20\% Gaussian bandwidth
centered at 90 GHz and $15^{\circ}$ FWHM horns. Approximate
kernel: full coloured line.  Refined kernel numerically computed: dotted
coloured line.  Kernel for monochromatic visibilities: black full line.
Right: same but with a top hat shaped bandwidth.}

\label{fig:exactvsappkernel} 
\end{figure*}

In the following, to allow for complete analytic calculation, we first
ignore the frequency dependence of the signal and of the beam. We
write: 
\begin{equation}
\tilde{A}(\vec{w})=\tilde{A}_{\nu_{0}}(\vec{w})\approx\tilde{A}_{\nu}(\vec{w})\ \ \mathrm{and}\ \ \tilde{I}(\vec{w})=\tilde{I}_{\nu_{0}}(\vec{w})\approx\tilde{I}_{\nu}(\vec{w}).
\end{equation}
This defines the approximate form of the convolution kernel: 
\begin{equation}
\tilde{\beta}_{ap}(\vec{u_{0}},\vec{w})=\int\tilde{A}(\vec{u_{0}}\frac{\nu}{\nu_{0}}-\vec{w})\ J_{N}(\nu-\nu_{0})\ d\nu.\label{eq:convkernap}
\end{equation}
This approximation allows us to get an intuitive idea of how the
bandwidth smearing acts with good enough accuracy to estimate the
sensitivity loss. We discuss in subsection \ref{sub:RefKernel} a
refined form of the kernel that takes into account the frequency dependence
of the beam and the intensity. As shown in figure \ref{fig:exactvsappkernel},
the difference between the approximate kernel, derived in subsection
1.3, and the refined kernel is small.

\begin{figure*}[!ht]
 \resizebox{\hsize}{!}{\centering{ \includegraphics{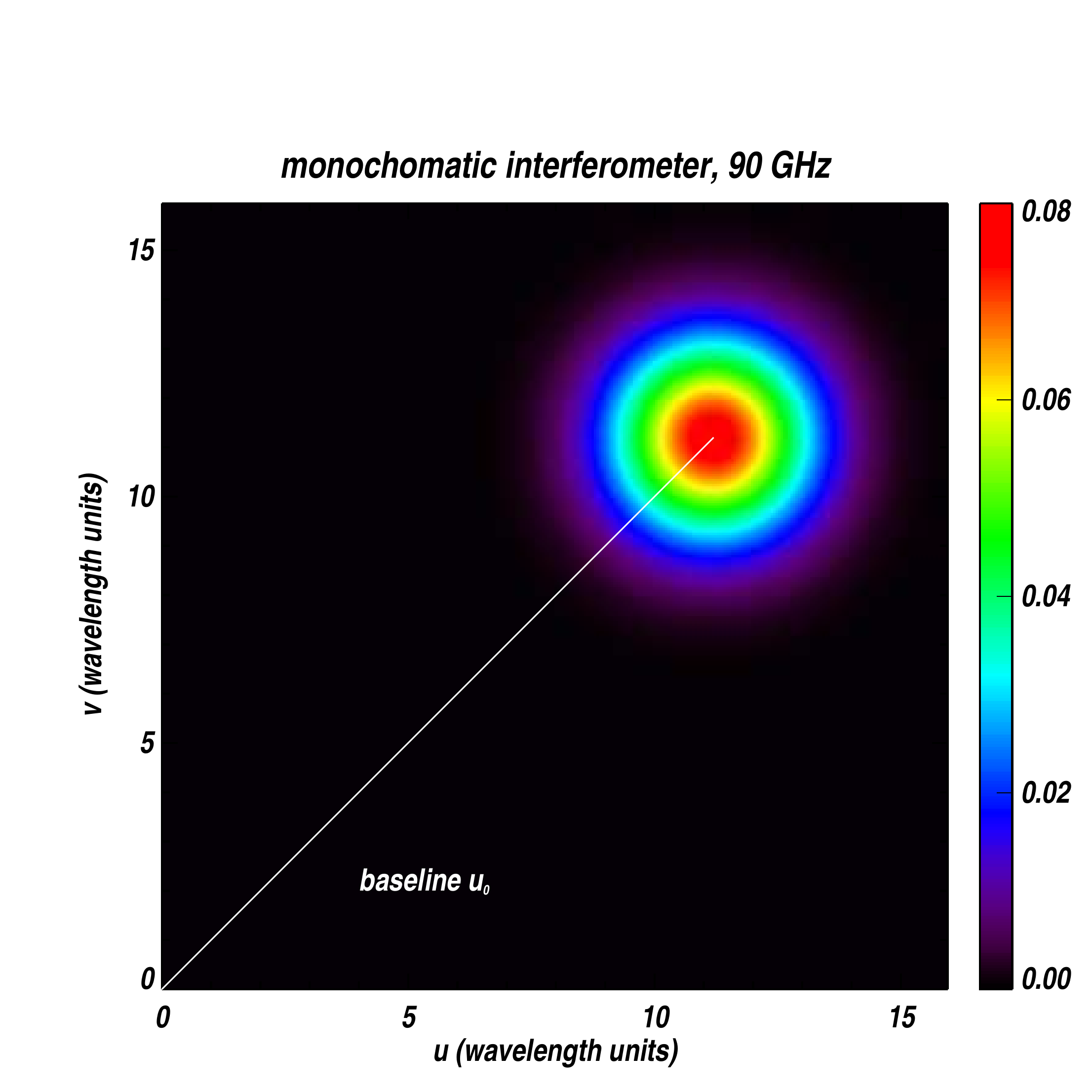}\includegraphics{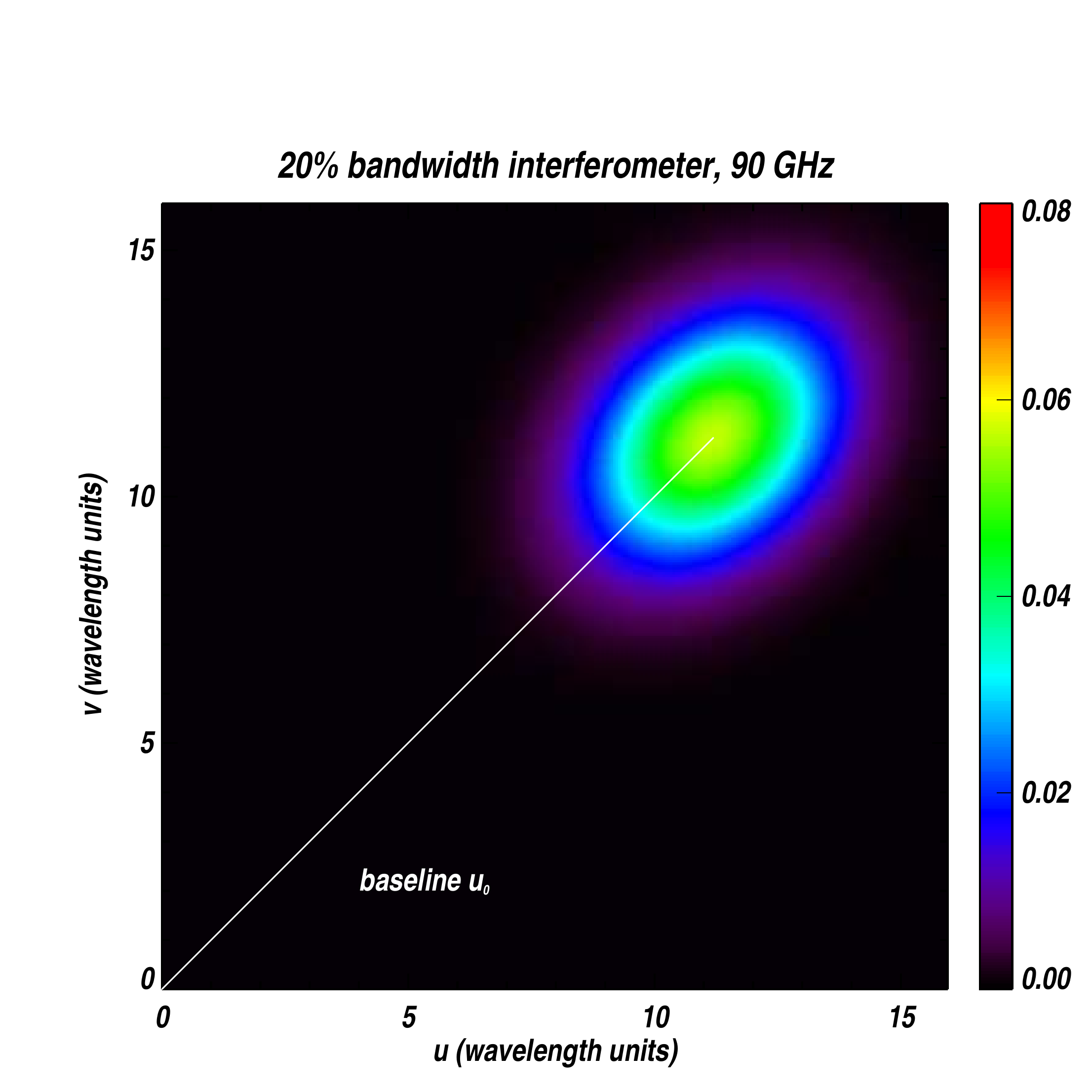}
}} 

\caption{Left: convolution kernel in the uv-plane for a monochromatic
interferometer, which is actually just the Fourier transform of the
primary beam, for a 90 GHz central frequency and a gaussian beam with
$15^{\circ}$ FWHM. Right: approximate convolution kernel in the
uv-plane for a baseline $\vec{u_{0}}$ (matching $l\sim100$), for a
20\% bandwidth interferometer, for a gaussian bandpass function
centered at 90 GHz and a gaussian beam with $15^{\circ}$ FWHM.}

\label{fig:convkernel} 
\end{figure*}

\subsection{Approximate analytical form of the kernel}

In order to carry out the analytical calculation, we also assume a
gaussian normalized bandpass function : \begin{equation}
J_{N}(\nu-\nu_0)=\frac{1}{\sigma_{\nu}\sqrt{2\pi}}\exp\left(-\frac{(\nu-\nu_0)^{2}}{2\sigma_{\nu}^{2}}\right).\label{eq:gaussbp}\end{equation}
The instrument bandwidth is related to the gaussian sigma by $\Delta\nu=\sigma_{\nu}\sqrt{2\pi}$.
We assume a gaussian beam for the horns, with the usual convention
$A(0)=1$, leading to the following beam in the uv-plane: \begin{equation}
\tilde{A}(\vec{w})=\Omega\exp(-\pi\Omega\vec{w}^{2}).\label{eq:gaussianbeam} \end{equation}
 As previously explained, we ignore here the frequency dependence
of the beam. The integral of the beam over the sky is defined for
the central frequency, $\Omega=\Omega_{\nu_{0}}$%
\footnote{This solid angle is then related to the RMS of the gaussian beam $\sigma$
by $\Omega=2\pi\sigma^{2}$.%
}. The expression of the kernel is then: 
\begin{equation}
\tilde{\beta}_{ap}(\vec{u_{0}},\vec{w})=\frac{\Omega}{\sigma_{\nu}\sqrt{2\pi}}\int\exp{\left[-\pi\Omega(\vec{u_{0}}\frac{\nu}{\nu_{0}}-\vec{w})^{2}-\frac{(\nu-\nu_{0})^{2}}{2\sigma_{\nu}^{2}}\right]}\ d\nu.\label{eq:firstbeta}
\end{equation}
This can be analytically integrated (details are given in appendix~\ref{app:betakernel})
and written in the form: 
\begin{equation}
\tilde{\beta}_{ap}(\vec{u_{0}},\vec{w'})=\tilde{A}(\vec{w'})\ \frac{\mathrm{exp}\left[\frac{2\pi^{2}\Omega^{2}\left(\frac{\sigma_{\nu}}{\nu_{0}}\right)^{2}}{1+2\pi\Omega\left(\frac{\sigma_{\nu}}{\nu_{0}}\right)^{2}\vec{u_{0}}^{2}}(\vec{u_{0}}\cdot\vec{w'})^{2}\right]}{\sqrt{1+2\pi\Omega\left(\frac{\sigma_{\nu}}{\nu_{0}}\right)^{2}\vec{u_{0}}^{2}}},
\label{eq:secondbeta}\end{equation}
where we have made the variable substitution
\begin{equation}
\vec{w'}=\vec{u_{0}}-\vec{w}.
\end{equation}
We can define the \textit{effective beam} in real space for a
broadband interferometer $\Omega_{s}$ as the value at $\vec{w'}=\vec{0}$
of the Fourier transform of the kernel\footnote{Because $J_N$ is normalized to 1, the inverse transform of $\tilde{\beta}_{ap}$ equals 1 at the top of the beam.}: \begin{equation}
\Omega_{s}=\tilde{\beta}_{ap}(\vec{u_{0}},\vec{0})=\frac{\Omega}{\kappa_{1}}.\label{eq:effbeam}\end{equation}
 Here \begin{equation}
\kappa_{1}=\sqrt{1+2\pi\Omega\left(\frac{\sigma_{\nu}}{\nu_{0}}\right)^{2}\vec{u_{0}}^{2}}\ ,\ \ \ \ (\kappa_{1}\geq1)\label{eq:kappa1def}\end{equation}
 We can finally rewrite the kernel: 
\begin{eqnarray}
\tilde{\beta}_{ap}(\vec{u_{0}},\vec{w'}) & = & \tilde{A}(\vec{w'})\times\frac{1}{\kappa_{1}}\times\exp\left[2\pi^{2}\left(\frac{\sigma_{\nu}}{\nu_{0}}\right)^{2}\Omega_{s}^{2}(\vec{u_{0}}\cdot\vec{w'})^{2}\right]\label{eq:thirdbeta}\\
 & = & \tilde{A}(\vec{w'})\times\frac{1}{\kappa_{1}}\times\exp\left[\pi\Omega\left(1-\frac{1}{\kappa_{1}^{2}}\right)\frac{(\vec{u_{0}}\cdot\vec{w'})^{2}}{\vec{u_{0}}^{2}}\right].
\label{eq:fourthbeta}\end{eqnarray}

 This $\kappa_{1}$ factor is an indicator of the importance of
bandwidth smearing. It reaches its minimal value 1 for a monochromatic
interferometer. One immediately sees that interferometers with ``small''
primary beams and/or ``small'' baselines
are less affected by the bandwidth smearing (see figure~\ref{fig:spf}).
We compare the convolution kernels of monochromatic
and broadband interferometers in figure~\ref{fig:convkernel}. One
can clearly see that the effect of the bandwidth smearing is to stretch
the kernel, in the baseline direction only, by which the Fourier transform
of the signal is convolved. When $\kappa_{1}$ is close to 1 (i.e.
$\left(\frac{\sigma_{\nu}}{\nu_{0}}\right)^{2}\vec{u_{0}}^{2}\ll\frac{1}{2\pi\Omega}$),
the size of the ``bandwidth part'' of the
kernel is smaller than the ``beam part''
and thus the signal is not degraded by the bandwidth: this is quite
intuitive since a signal already convolved by a kernel of width $\sigma_{a}$
is not significantly
degraded if it is convolved again by a kernel of width $\sigma_{b}$
such as $\sigma_{b}<\sigma_{a}$. The physical interpretation in
\textit{real space} is that the bandwidth smearing makes the beam
narrower, by a factor $\kappa_{1}$ that depends on the baseline length:
for a given multipole $l$, the fraction of sky observed by a broadband
interferometer is actually $f_{sky}^{\Delta_{\nu}}(l)=f_{sky}(l)/\kappa_{1}$
-- cf. section~\ref{sec:sensitivity}.

\subsection{Broadband visibilities in temperature units for CMB experiments}

It is more convenient to work with visibilities in temperature units
when studying CMB temperature and polarization anisotropies%
\footnote{Practically, the visibilities measured by a bolometric interferometer
will be in power units as defined in Eq.~(\ref{eq:broadvispower}).%
}. If $B_{\nu}$ is the intensity of the observed field, in units of
{[}$\mathrm{W.Hz^{-1}.m^{-2}.sr^{-1}}${]}, the spectral power collected
by a horn of surface $S$ is: 
\begin{equation}
I_{\nu}=SB_{\nu}.\label{eq:specpower}
\end{equation}
CMB experiments observe small spatial fluctuations over the sky:
\begin{equation}
I_{\nu}(\vec{n})=I_{\nu}+\Delta I_{\nu}(\vec{n}).
\end{equation}
The oscillating term of the visibilities washes out the constant
part of the spectral power, so the visibilities can be rewritten
\begin{equation}
V_{I}^{\Delta\nu}(\vec{u_{0}})=\iint\Delta I_{\nu}(\vec{n})A_{\nu}(\vec{n})\ \mathrm{exp}(i2\pi\vec{d_{0}}\cdot\vec{n}\frac{\nu}{c})\ J(\nu-\nu_{0})\ d\nu\ d\vec{n}\end{equation}
The temperature fluctuations over the sky are linked to the power
fluctuations by
\begin{equation}
\Delta I_{\nu}(\vec{n})=\frac{\partial I_{\nu}}{\partial T}\Delta T(\vec{n})=S\frac{\partial B_{\nu}}{\partial T}\Delta T(\vec{n}).
\end{equation}
We can then define the broadband visibilities in temperature units:
\begin{equation}
V_{T}^{\Delta\nu}(\vec{u_{0}})[\mathrm{in\
K]}=\frac{V_{I}^{\Delta\nu}(\vec{u_{0}})[\mathrm{in\
W]}}{S\left.(\partial B_{\nu}/\partial T)\right|_{\nu_{0}}}.
\label{eq:tempvisdef}\end{equation}
 Following the same arguments as previously, one can show that
\begin{equation}
V_{T}^{\Delta\nu}(\vec{u_{0}})=\Delta\nu\int\tilde{\Delta T}(\vec{w})\tilde{\beta}^{T}(\vec{u_{0}},\vec{w})\ d\vec{w},
\end{equation}
where we have introduced the Fourier transform of the temperature
field,
\begin{equation}
\Delta T(\vec{n})=\int\tilde{\Delta T}(\vec{w})\exp(-i2\pi\vec{w}\cdot
\vec{n})d\vec{w},
\end{equation}
and a \textit{temperature convolution kernel}, 
\begin{equation}
%
\tilde{\beta}^{T}(\vec{u_{0}},\vec{w})=\int\left(
\frac{
\partial
B_{\nu}/\partial T}{\left.(\partial B_{\nu}/\partial
T)\right|_{\nu_{0}}}\right)\tilde{A_{\nu}}(\vec{u_{0}}\frac{\nu}{\nu_{0}}-\vec{w})\
J_{N}(\nu-\nu_{0})\ d\nu.
\label{eq:convkertemp}
\end{equation}
If we neglect the dependence of the signal and of the beam on frequency,
this kernel actually becomes the one defined in Eq.~(\ref{eq:convkernap}):
\begin{equation}
\tilde{\beta}_{ap}^{T}(\vec{u_{0}},\vec{w})\rightarrow\tilde{\beta}_{ap}(\vec{u_{0}},\vec{w})\end{equation}

\begin{figure*}[!t]
\begin{centering}
\includegraphics{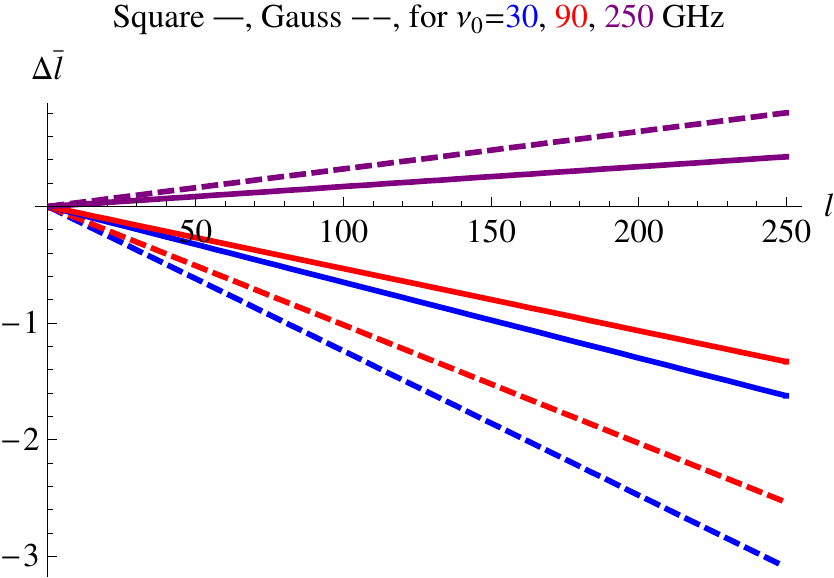}\includegraphics{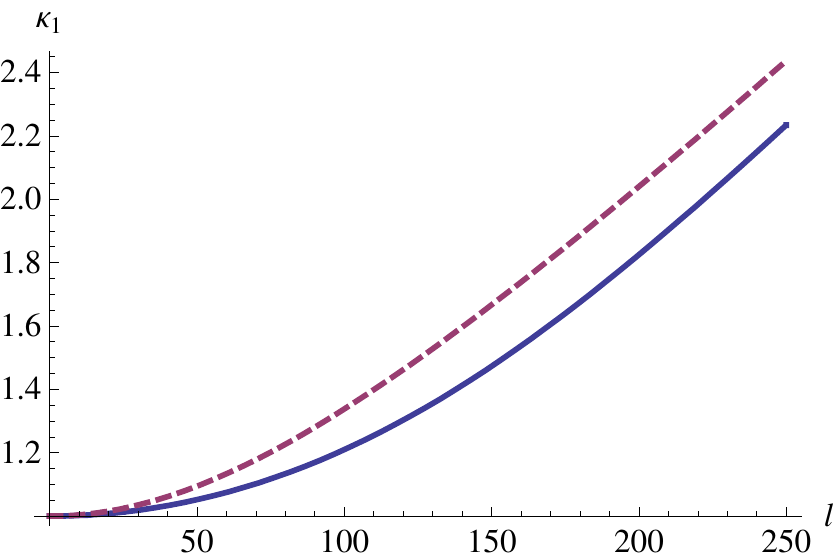}
\par\end{centering}

\caption{Left: Shift of the effective $l$ window function as a function of
$l_{0}$; dashed lines for a gaussian bandwidth, solid lines for a
top hat one.\protect \\
Right: The variation of $\kappa_{1}$ as a function of $l_{0}$;
dashed lines for a gaussian bandwidth, solid lines for a top hat one.\label{fig:lshift-kappa}}

\end{figure*}

\subsection{A refined kernel\label{sub:RefKernel}}

The intensity $B_{\nu}$ of the observed field actually depends on
frequency for a black body source at temperature $T$: 
\begin{equation}
B_{\nu}=\frac{2h\nu^{3}}{c^{2}}\frac{1}{\exp\left({h\nu}/{k_{B}T}\right)-1} .
\label{eq:bbintensity}
\end{equation}
Inside the bandwidth, the frequency dependence of the $T$ derivative
of $B_{\nu}$ is well approximated by a power law : 
\begin{equation}
\frac{\partial B_{\nu}}{\partial T}\simeq\left.\frac{\partial
B_{\nu}}{\partial
T}\right|_{\nu_{0}}\left(\frac{\nu_{0}}{\nu}\right)^{\alpha}\mbox{
with
}\alpha=\frac{h\nu_{0}}{k_{B}T}\,\frac{e^{{h\nu_{0}}/{k_{B}T}}+1}{e^{{h\nu_{0}}/{k_{B}T}}-1}.
\label{eq:dpdt}\end{equation}
The beam of the horns depends on frequency as well. The surface of
the horns $S$, the solid angle covered $\Omega$ and the frequency
of observation $\nu$ are related by $S\Omega=\kappa\frac{c^{2}}{\nu^{2}}$,
leading to $\Omega_{\nu}=\Omega_{\nu_{0}}\frac{\nu_{0}^{2}}{\nu^{2}}$.
Such a frequency dependence of a Gaussian beam can be modeled by
modifying its Fourier transform as follows: \begin{equation}
\tilde{A_{\nu}}(\vec{w})=\Omega\frac{\nu_{0}^{2}}{\nu^{2}}\exp(-\pi\Omega\frac{\nu_{0}^{2}}{\nu^{2}}\vec{w}^{2}).\label{eq:BeamDep}\end{equation}
Figure~\ref{fig:exactvsappkernel} shows for several values of $l$
how the approximate kernel of Eq.~(\ref{eq:fourthbeta}) is refined
when one takes into account the above frequency dependences of the
beam and the intensity.
The signal and the beam dependencies largely
compensate each other, and the difference between the two kernels
turns out to be negligible considering the accuracy level required for the
sensitivity loss estimation. The main difference is a shift of the
centroid of the window function shown in the left panel of figure
\ref{fig:lshift-kappa}, absent in the approximate kernel. This will
introduce systematics which will have to be corrected for. We also
show what happens for a more realistic top hat shaped bandwidth (right
pannel of figure \ref{fig:exactvsappkernel}, solid lines in figure
\ref{fig:lshift-kappa}). For the data extraction of a given instrument,
the convolution kernel will have to be computed numerically; however,
the approximate Gaussian kernel gives a good enough accuracy to estimate
the sensitivity loss. In the rest of this paper, we use the approximate
analytical form of the kernel, and write $\tilde{\beta}$ instead
of $\tilde{\beta}_{ap}$. Finally, the right panel of figure \ref{fig:lshift-kappa}
shows the variation of $\kappa_{1}$ with $l_{0}$ for a gaussian
(dashed line) and a top hat (solid line) bandwidth.
 
\section{Visibilities measured by broadband bolometric interferometers \label{sec:broadbolvis}}
The effect of the bandwidth is more subtle in a bolometric
interferometer than in a generic interferometer, because the visibilities
are not measured \textit{directly}. As described in [C08], a time-domain modulation of the visibilities is performed by controlled phase shifters -located behind each polarization channel (twice the number of horns)- which take some well-chosen time-sequences of discrete phase values. The corresponding time-sequences of bolometers' measurements will allow to recover, independently for each bolometer, all the different visibilities, by solving a linear problem of the form $S=A \cdot X$, where
$X$ is a vector including the visibilities, $S$ is a vector including a time-sequence
of one bolometer's measurements, and $A$ a coefficient matrix depending on the
phase shift sequences. In the following we generalize the [C08]
formalism, taking the bandwidth into account. As in the generic case,
we assume in this section that the detectors (here the bolometers) are sensitive to a
spectral bandwidth $\Delta \nu$, through a bandpass function $J(\nu-\nu_0)$
centered on the frequency $\nu_0$.

\subsection{Signal on broadband bolometers}
We consider a bolometric interferometer consisting of $N_h$ horns
whose beams are defined by the same function $B(\vec{n})$, and
$N_{out}$ bolometers. The electric field at the output of polarization
splitters, corresponding to horn $i$ coming from direction $\vec{n}$
for polarization $\eta \ (\| \mathrm{\ or \ } \bot)$ is
\begin{equation}
\epsilon_{i,\eta}(\vec{n}) = B(\vec{n}) E_{i,\eta}(\vec{n}).
\end{equation}
During a time sample $k$, each controlled phase shifter adds to its associated input channel the phase $\Phi^k_{i,\eta}(\nu)$. After combining, the electric field in one output channel is then
\begin{equation}
z_k(\vec{n}) = \int {z_k^\nu(\vec{n})} J(\nu-\nu_0) d\nu,
\end{equation}
where
\begin{equation}
z_{k}^\nu(\vec{n})=\frac{1}{\sqrt{N_{\mathrm{out}}}}\sum_{i=0}^{N_{h}-1}\sum_{\eta=0}^{1}\epsilon_{i,\eta}(\vec{n})\exp[i\Phi^k_{i, \eta}(\nu)],  
\end{equation}

The power coming from each of the combiner outputs is averaged on time
scales given by the time constant of the detector (much larger than the
frequency of the EM wave). The power collected by a given bolometer during a time sample $k$\footnote{Recall that this is a sequence of such time samples that will be used to inverse the problem and recover the visibilities.} is then
\begin{equation}
S_k=\left<\left| \iint{z^{\nu}_k(\vec{n}) J(\nu-\nu_0) d\nu d\vec{n}} \ \right|^2 \right>_{time}.
\end{equation}
Signals coming from different directions of the sky are
incoherent, as are signals at different frequencies, so
their time averaged correlations vanish:
\begin{equation}
\left<z_{k}^{\nu}(\vec{n})z_{k}^{\nu^{\prime} \star}(\vec{n}^{\prime})\right>_{\mathrm{time}}  = \left|z^{\nu}_{k}(\vec{n})\right|^{2}\delta(\vec{n}-\vec{n}^{\prime}) \ \delta(\nu - \nu^{\prime})
 \end{equation}
The signal on the bolometer is finally:
\begin{equation}
S_k = \iint{ \vert z^{\nu}_k(\vec{n}) \vert ^2 J(\nu - \nu_0) d\nu d\vec{n} }
\end{equation}
Developing this expression leads to auto-correlation terms for each input channel and cross-correlation terms between all the possible pairs: 
\begin{equation}
S_k = S_k^{auto} \ + \ S_k^{cross} 
\end{equation}
As in the general case, one can write the visibilities as a convolution in the uv-plane (we define $b=\{i,j\}$ and $p=\{\eta_1, \eta_2\}$): 
\begin{equation}
S_{k}^{cross} = \frac{2 J(0) \ \Delta \nu}{N_{out}} \mathrm{Re}\big[ \sum_{i<j} \sum_{\eta_1, \eta_2} \int \tilde{I}_{b,p}(\vec{w}) \tilde{\eta}^{BI}_{b,p}(\vec{u_b}, \vec{w}) d\vec{w} \big] \label{eq:bolosignal1}
\end{equation}
where we have defined the Fourier transform of the physical signal,
\begin{equation}
\tilde{I}_{b,p}(\vec{w}) = \int E_{i, \eta_1}(\vec{n})E^\star_{j,\eta_2}(\vec{n}) \exp(i2\pi\vec{n}.\vec{w}) d\vec{n} \ ,
\end{equation}
and the kernel $\tilde{\eta}^{BI}_{b,p}(\vec{u_b}, \vec{w})$ which contains the phase modulation used to recover the visibilities,
\begin{eqnarray}
  \tilde{\eta}^{BI}_{b,p}(\vec{u_b}, \vec{w}) &=&  \int \tilde{A}(\frac{\nu}{\nu_0} \vec{u_b} - \vec{w}) \ J_N(\nu-\nu_0) \ \exp[i\Delta \Phi^ k_{b,p}(\nu)] \ d\nu \label{eq:firsteta} \\
  \Delta \Phi^k_{b,p}(\nu) &=& \Phi^k_{i,\eta_1}(\nu) - \Phi^k_{j, \eta_2}(\nu)
\end{eqnarray}
with $A(\vec{n})=B^2(\vec{n})$ the square of the beam of the input horns.
\subsection{Phase shifters constant with respect to frequency}
We first consider the simplest case where the phase shift values do not depend on the frequency $\nu$. When $\Delta \Phi^k_{b,p}(\nu) = \Delta \Phi^k_{b,p}(\nu_0) $, the phase shift term comes out the integral over $\nu$: 
\begin{equation}
 \tilde{\eta}^{BI}_{b,p}(\vec{u_b}, \vec{w}) = \tilde{\beta}(\vec{u_b}, \vec{w}) \exp[i\Delta \Phi^k_{b,p}(\nu_0)]. \label{eq:etakernana}
\end{equation}
The signal of the cross correlations on the bolometer is thus the one expressed in [C08], with the broadband visibilities defined in Eq.~(\ref{eq:tempvisdef}) instead of the monochromatic ones:
 \begin{eqnarray}
 S_{k}^{cross} &=& \frac{2}{N_{out}} \mathrm{Re}\big[ \sum_{i<j} \sum_{\eta_1, \eta_2} e^{i\Delta \Phi^k_{b,p}(\nu_0)} V^{\Delta_\nu}_{b,p}(\vec{u_b}) \big].  \label{eq:scross1}
\end{eqnarray} 
Following [C08], we can introduce the \textit{broadband Stokes visibilities},
\begin{equation}
V^{\Delta_\nu}_{S}(\vec{u_b}) = J(0) \ \Delta \nu \int \tilde{S}(\vec{w}) \tilde{\beta}(\vec{u_b}, \vec{w}) \ d\vec{w},
\end{equation}
where $S$ stands for the Stokes parameters I, Q, U or V and $\tilde{S}$ stands for their Fourier transform, and rewrite Eq.~(\ref{eq:scross1}) as a linear combination of the Stokes visibilities defined for different baselines: 
\begin{eqnarray}
  S_{k}^{cross} &=& \sum_{\beta=0}^{N_{\neq}-1} \Gamma_{k, \beta} \cdot \mathsf{V}^{\Delta_\nu}_{\beta} \label{eq:linprobfps1}
\end{eqnarray}
where $\Gamma_{k, \beta}$ is the vector, defined in [C08], encoding the phase shifting values, and $\mathsf{V}^{\Delta_\nu}_{\beta}$ is a vector including the real and imaginary parts of the broadband Stokes visibilities. If the phase shift values of a broadband bolometric interferometer are constant with respect to frequency, the visibilities should thus be reconstructed exactly as explained in [C08], by solving a linear problem. In this case, a broadband bolometric interferometer therefore works exactly as a monochromatic one, except that the output observables will be broadband visibilities instead of monochromatic ones. 

\subsection{Phase shifters linear with respect to frequency \label{subsec:fdps}}
We now consider the more complicated case where the modulation phase shifters vary linearly with respect to frequency:
\begin{equation}
\Delta \Phi^k_{b,p}(\nu) = \Delta \Phi^k_{b,p}(\nu_0) \times \frac{\nu}{\nu_0}.
\end{equation}
From a technological point of view, this may seem more natural, since it is automatically respected if for instance the phase shifters are just constituted by delay-lines.

For the sake of simplicity, we write $ \Delta\Phi^k_{b,p}$ instead of
$ \Delta\Phi^k_{b,p}(\nu_0)$ in the following.  As in
section~\ref{sec:broadvis}, in order to carry out the analytical
calculation, we assume both the beam and the intensity 
to be independent of frequency,
and we assume a
normalized gaussian bandpass function $J_N$. We show in
appendix~\ref{app:etakernel} that $\tilde{\eta}^{BI}_{b,p}$ is then
 \begin{equation}
  \tilde{\eta}^{BI}_{b,p}(\vec{u_b}, \vec{w}) =  \tilde{\beta}(\vec{u_{b}}, \vec{w}) \ \exp{\left[-(\Delta\Phi_{b,p}^{k})^2 G + i \Delta\Phi^k_{b,p} \left(1 \ - \ H(\vec{w})\right) \right]}, \label{eq:secondeta}
\end{equation}
where we have defined\footnote{$\kappa_1$ is defined in Eq.~(\ref{eq:kappa1def}).}
\begin{eqnarray}
G &=& \left(\frac{\sigma_\nu}{\nu_0}\right)^2 \times \frac{1}{2\kappa_1^2}, \label{eq:G}\\
H(\vec{w}) &=& \left(1 - \frac{1}{\kappa_1^2}\right) \times \frac{\vec{u_b} \cdot (\vec{u_b}-\vec{w})}{\vec{u_b}^2}.
\end{eqnarray}
 Thus the cross-correlation part of the bolometer signal can be written:
\begin{eqnarray}
 S_{k}^{cross} &=& \frac{2}{N_{out}} \mathrm{Re}\big[ \sum_{i<j} \sum_{\eta_1, \eta_2}  e^{-(\Delta\Phi^k_{b,p})^2 G \ +\  i\Delta \Phi^k_{b,p}} \ V^{LD, k, \Delta_\nu}_{b,p}(\vec{u_b}) \big] \label{eq:crossbi}
\end{eqnarray}
where we have introduced some "phase-dependent" broadband visibilities:
\begin{equation}
V^{LD, k, \Delta_\nu}_{b,p}(\vec{u_b}) = J(0) \ \Delta \nu \int \tilde{I}_{b,p}(\vec{w}) \tilde{\beta}_{b,p}^{LD, k}(\vec{u_b}, \vec{w}) \ d\vec{w}
\end{equation}
The new kernel is linked to the generic one by a rotation in the complex plane:
\begin{equation}
\tilde{\beta}^{LD, k}_{b,p}(\vec{u_b}, \vec{w}) = \tilde{\beta}(\vec{u_b}, \vec{w}) \exp \left[-i \Delta\Phi^k_{b,p} H(\vec{w}) \right] \label{eq:cplexfactor}
\end{equation}

This complex factor unfortunately depends on the phase differences: this means that the definition of every visibility will slightly change between two different samples $k$ and $k'$! This is of course a bad feature that will corrupt the linear problem. 

We will show in a paper in preparation that an error varying with the
modulation will lead to a dramatic leakage from the intensity
visibilities into the polarization ones (which are at least two orders
of magnitudes smaller in CMB observations). This prediction (which is
not trivial and is not proven analytically in this article) is
supported by our Monte-Carlo simulation (cf. section
\ref{sec:mcsimu}): introduction of the $\eta$-kernel of
Eq.~(\ref{eq:secondeta}) in the simulation leads to a huge error on
the reconstructed polarization spectrum (typically two orders of
magnitude bigger than the one on temperature spectrum), as shown in
figure~\ref{fig:recspec_fps}, when the modulation matrix used to solve
the problem is the monochromatic one defined in [C08].  Fortunately,
there is a way to get rid of this leakage, as described in section
\ref{sec:virtsub}, by reconstructing the $I$ visibilities in
sub-bands. Using the extended modulation matrix introduced in
subsection~\ref{subsec:spescheme}, this dramatic error source can be
put under control, and the broadband polarization visibilities can be
reconstructed without loss of sensitivity.

\subsection{Geometrical phase shifts}
In the quasi-optical combiner design considered for the QUBIC experiment [\cite{qubicjean}], some geometrical phase shifts $\Psi^k_{b,p}(\nu)$ are automatically introduced by the combiner\footnote{As mentioned in [C08], these phase shifters naturally respect the "coherent summation of equivalent baselines" scheme.}. These phase shifts, stemming from path differences between rays in the optical combiner, will vary linearly with respect to frequency:
\begin{equation}
\Delta \Psi^k_{b,p}(\nu) = \Delta \Psi^k_{b,p}(\nu_0) \times \frac{\nu}{\nu_0}.
\end{equation}
 But as explained in [C08], these geometrical phase shifts will not be used to modulate the visibilities\footnote{As geometrical phase shifts depend on the spatial positions of bolometers in the quasi-optical combiner focal plane, using them to invert the problem means using different bolometers; this must be avoided because of dangerous intercalibration issues.}. This means that they will not vary between the different time samples used for a reconstruction:
 \begin{equation}
 \forall k, k', \ \ \ \Delta \Psi^k_{b,p}(\nu) = \Delta \Psi^{k'}_{b,p}(\nu) \equiv \Delta \Psi_{b,p}(\nu) 
 \end{equation}
Hence they will not cause any error during the reconstruction: the $\beta$-kernels of each different visibilities will be rotated in the complex plane by the same factor whatever the sample. The visibilities will just be measured with a rotation:
\begin{equation}
V^{GLD, \Delta_\nu}_{b,p}(\vec{u_b}) = J(0) \ \Delta \nu \int \tilde{I}_{b,p}(\vec{w}) \tilde{\beta}(\vec{u_b}, \vec{w}) \exp \left[-i \Delta\Psi_{b,p} H(\vec{w}) \right]  \ d\vec{w}.
\end{equation}

\subsection{Photon noise error on reconstructed visibilities}
We suppose here that the modulation matrix used to reconstruct the
visibilities is the monochromatic one defined in [C08]; this is
completely true in the case of frequency-independent phase shifters,
and true for the rows regarding the polarization visibilities in the
case of frequency-dependent phase shifters -- see
subsection~\ref{subsec:spescheme}. We have shown in [C08] that the
visibility covariance matrix is (the factor $1/[J(0)\ \Delta
\nu]^2$ comes from the difference of definition between monochromatic
and broadband visibilities)
\begin{equation}
N=\frac{\sigma_0^2 N_h}{\left[J(0) \ \Delta \nu \right]^2 \ N_{out}} \times \left( A^t \cdot A \right)^ {-1},
\end{equation}
where $A$ is a matrix including the $\Gamma_{k, \beta}$
vectors and $\sigma_0$ is the quantity of photon noise (in watts) that
would be seen by one bolometer illuminated by one horn during the time
of one sample of the phase sequence. The off-diagonal
elements cancel out to zero because the angles are uncorrelated from
one channel to another, while the diagonal elements average to the
variance of the elements in $A$, which equals 1\footnote{$A$ is filled
with elements of the form $\cos(\phi_1) + \sin(\phi_2)$ and because we
are assuming that the angles are uniformly distributed, they have zero
average and their variance is 1.}, multiplied by the number of
different data samples $N_d$. If the coherent summation of equivalent
baselines scheme is adopted, the variance on the reconstructed
visibilities is
\begin{equation}
\sigma_{V}[\mathrm{in \ W}] = \frac{\sqrt{N_h}}{N_{eq}} \frac{\sigma_0}{\sqrt{N_t}} \times \frac{1}{J(0) \ \Delta \nu},  \label{eq:noisesigma0} 
\end{equation}
where $N_t=N_d \times N_{out}$ is the total number of time samples. We derive from Eq.~(\ref{eq:noisesigma0}) in appendix~\ref{app:noisek} the noise on a visibility measured during a time $t$ by a bolometric interferometer experiment, knowing the Noise Equivalent Temperature (NET\footnote{See appendix~\ref{app:net} for definition.}) of its bolometers, in Kelvin units:
\begin{equation}
\sigma_{V}[\mathrm{in \ K}] = \frac{\sqrt{N_h}}{N_{eq}} \frac{\mathrm{NET} \ \Omega}{\sqrt{t}}.  \label{eq:sigmavbi} 
\end{equation}

\section{Virtual reconstruction sub-bands in bolometric interferometry \label{sec:virtsub}}
In this section we show that the linear dependence in frequency of the
modulation phase shifts $\Phi^k_{b,p}(\nu)$ enables indepndent reconstruction
of the visibilities in smaller frequency sub-bands. This
idea has been proposed for the first time in [\cite{maluphd}]. We
first thought that this method could be a way to beat the smearing
(the sub-band visibilities that are reconstructed are less smeared
than the broadband ones), but we show in the following that its
application comes along with a loss in signal to noise ratio that
thwarts the gain in sensitivity, and thus makes this method
inneficient to beat bandwidth smearing. However, as we will see, this
method can be succesfully set up to remove the dramatic effect
described in subsection~\ref{subsec:fdps}, and thus saves the
frequency-dependent option for the modulation phase shifters.

\subsection{Principle}
Before doing the visibility reconstruction, one can choose a number
$n_{vsb}$ of \textit{virtual reconstruction sub-bands} of width
$\delta\nu=\Delta\nu/n_{vsb}$. We emphasize that this division in
sub-bands is purely virtual in that the hardware design does not
depend on it. The $\tilde{\eta}$-kernel becomes
\begin{equation}
 \tilde{\eta}^{BI, \Delta\nu}_{b,p}(\vec{u_b}, \vec{w}) = \sum_{m=1}^{n_{vsb}}  \tilde{\eta}^{BI, \delta\nu}_{b,p}[\vec{u_b}, \vec{w},  \nu_m],
\end{equation}
where $ \tilde{\eta}^{BI, \delta\nu}_{b,p}$ are defined as \textit{sub-bands kernels}:
\begin{equation}
 \tilde{\eta}^{BI, \delta\nu}_{b,p}[\vec{u_b}, \vec{w},  \nu_m] = \int \tilde{A}(\frac{\nu}{\nu_0} \vec{u_b} - \vec{w}) \ J_N^{\delta\nu}(\nu-\nu_m) \ \exp \left[i\Delta \Phi^k_{b,p}(\nu)\right] \ d\nu,
\end{equation}
where $J_N^{\delta\nu}(\nu-\nu_m)$ is a bandpass function of width
$\delta\nu$ centered at $\nu_m$. The cross-correlation term is then%
\footnote{We omit here the correction terms
involving $G$ and $H$ in order to simplify the expression in the case
of frequency-dependent phase shifters.}
\begin{equation}
  S_{k}^{cross} =  \frac{2}{N_{out}} \mathrm{Re}\left[ \sum_{m=1}^{n_{vsb}} \sum_{\substack{i<j \\ \eta_1, \eta_2}} \exp \left[ i\Delta \Phi^k_{b,p}  \right] V^{\delta\nu}_{b,p}(\vec{u_{b,m}}) \right], \label{eq:crossbivsb}
 \end{equation}
 where we have defined 
 \begin{equation}
 \Delta\Phi_{b,p,m} =  \Delta\Phi_{b,p}(\nu_0) \times \frac{\nu_m}{\nu_0} \ \ and \ \ \vec{u_{b,m}} = \vec{u_b} \times \frac{\nu_m}{\nu_0}.
 \end{equation}

The sub-band visibilities $V^{\delta\nu}_{b,p}(\vec{u_{b,m}})$
are the broadband visibilities defined in the previous sections, but
for a $\delta\nu$ bandwidth. For each pair of horns, the $n_{vsb}$
sub-bands visibilities are defined for $n_{vsb}$ different baselines
$\{u_{b,m}\}_{m=1,..., n_{vsb}}$.  Finally, Eq.~(\ref{eq:crossbivsb})
can be written as a linear system $n_{vsb}$ times greater than the one
of Eq.~(\ref{eq:crossbi}):
\begin{equation}
S^{cross}_{k} = \sum_{\beta=1}^{n_{vsb} \times N_{\neq}} \Gamma_{k, \beta}^{\delta\nu} \cdot \mathsf{V}^{\delta\nu}_{\beta}
\end{equation}
The problem can thus be inversed exactly as in [C08] to recover the $n_{vsb}\times N_{\neq}$ sub-bands visibilities.

\subsection{Sensitivity issue}
In every time sample $k$, the signal of the sub-band visibilities is $n_{vsb}$ times smaller than the signal of the broadband visibilities, and consequently their reconstruction variance is
\begin{equation}
\sigma^{\delta \nu}_{V}[\mathrm{in \ K}] = \frac{\sqrt{N_h}\times n_{vsb}}{N_{eq}  } \frac{\mathrm{NET} \ \Omega}{\sqrt{t}}.  \label{eq:sigmavbisubbands} 
\end{equation}

One can average \textit{offline} (i.e., after the reconstruction) the sub-band visibilities deriving from the same baseline $\vec{u_b}$, to recover the broadband visibilities:
\begin{equation}
\mathsf{V}^{\Delta\nu}_{b,p}(\vec{u_b})= \sum_{m=1}^{n_{vsb}} \mathsf{V}^{\delta\nu}_{b,p} (\vec{u_{b, m}}).
\end{equation}
These broadband visibilities are equivalent to the ones that would have been reconstructed without sub-bands division but suffer from a smaller bandwidth smearing ($\kappa_1$ is the one of an instrument whose bandwidth is $n_{vsb}$ times smaller). However, the variance on these broadband visibilities is
\begin{equation}
\sigma^{\Delta \nu}_{V}[\mathrm{in \ K}] = \frac{\sqrt{N_h}\times \sqrt{n_{vsb}}}{N_{eq}  } \frac{\mathrm{NET} \ \Omega}{\sqrt{t}}.  \label{eq:sigmavbisubbands2} 
\end{equation}
Comparison with Eq.~(\ref{eq:sigmavbi}) shows that the reconstruction in virtual sub-bands comes along with a loss of a factor $\sqrt{n_{vsb}}$ in sensitivity on reconstructed visibilities. Unfortunately, we will see in section \ref{sec:sensitivity} that this loss in sensitivity is always bigger than the gain due to the smearing reduction. This makes this method inefficient to beat bandwidth smearing.


\subsection{Reconstruction scheme for instruments with frequency-dependent phase shifters \label{subsec:spescheme}}
This method, however, happens to bring a solution to the crucial issue
described in subsection~\ref{subsec:fdps}. The idea is to estimate at
the same time the intensity visibilities in sub-bands (Stokes I), and
the polarization visibilities in one single broad band (Stokes Q,U and
V). This can easily be done by writing an extended coefficient matrix
based on the following decomposition:
\begin{equation}
S^{cross}_{k} = \sum_{\beta=1}^{n_{vsb} \times N_{\neq}} \Gamma_{k, \beta}^{\delta\nu} \cdot \mathsf{V}^{\delta\nu}_{I, \beta} + \sum_{X=Q,U,V} \sum_{\beta'=1}^{N_{\neq}} \Gamma_{k, \beta'} \cdot \mathsf{V}^{\Delta\nu}_{X, \beta'}.
\end{equation}
Practically, this means that the part of the matrix encoding the
polarization visibilities is identical to that of the monochromatic
matrix, while the part encoding the intensity visibilities gets
$n_{vsb}$ times more rows.  The matrix thus has a total of 
$(2 \times n_{vsb} +
6) \times N_{\neq}$ rows. The corruption of the linear problem (and
then the leakage of the error on the intensity visibilities into the
polarization ones) can thus be decreased as much as necessary by
increasing the number of sub-bands, without loss of signal to noise
ratio for the polarization visibilities. Such a reconstruction has
been performed with our numerical simulation, as described in
section~\ref{sec:mcsimu}; comparison of figures~\ref{fig:recspec_ldps}
and~\ref{fig:recspec_ldps_RVSB} shows its efficiency. One drawback of
this method is of course that it increases the minimal sequence length
required to invert the problem. Finally, it should be noticed that
this method could in principle apply just as well to phase shifters
with any arbitrary (but known) frequency dependence.

\begin{figure*}[!ht]
\centering\resizebox{\hsize}{!}{\centering{
\includegraphics[angle=0]{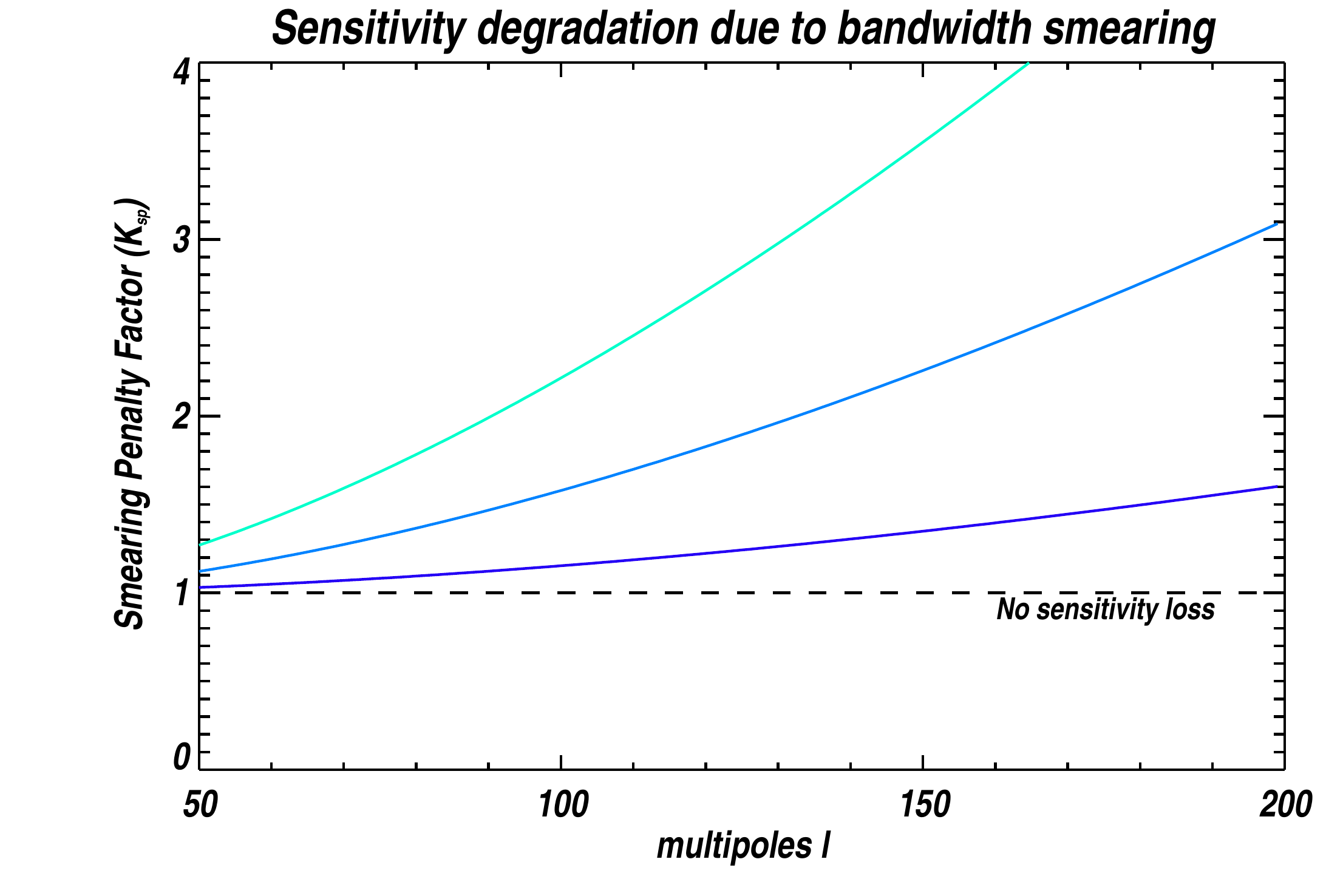} \includegraphics[angle=0]{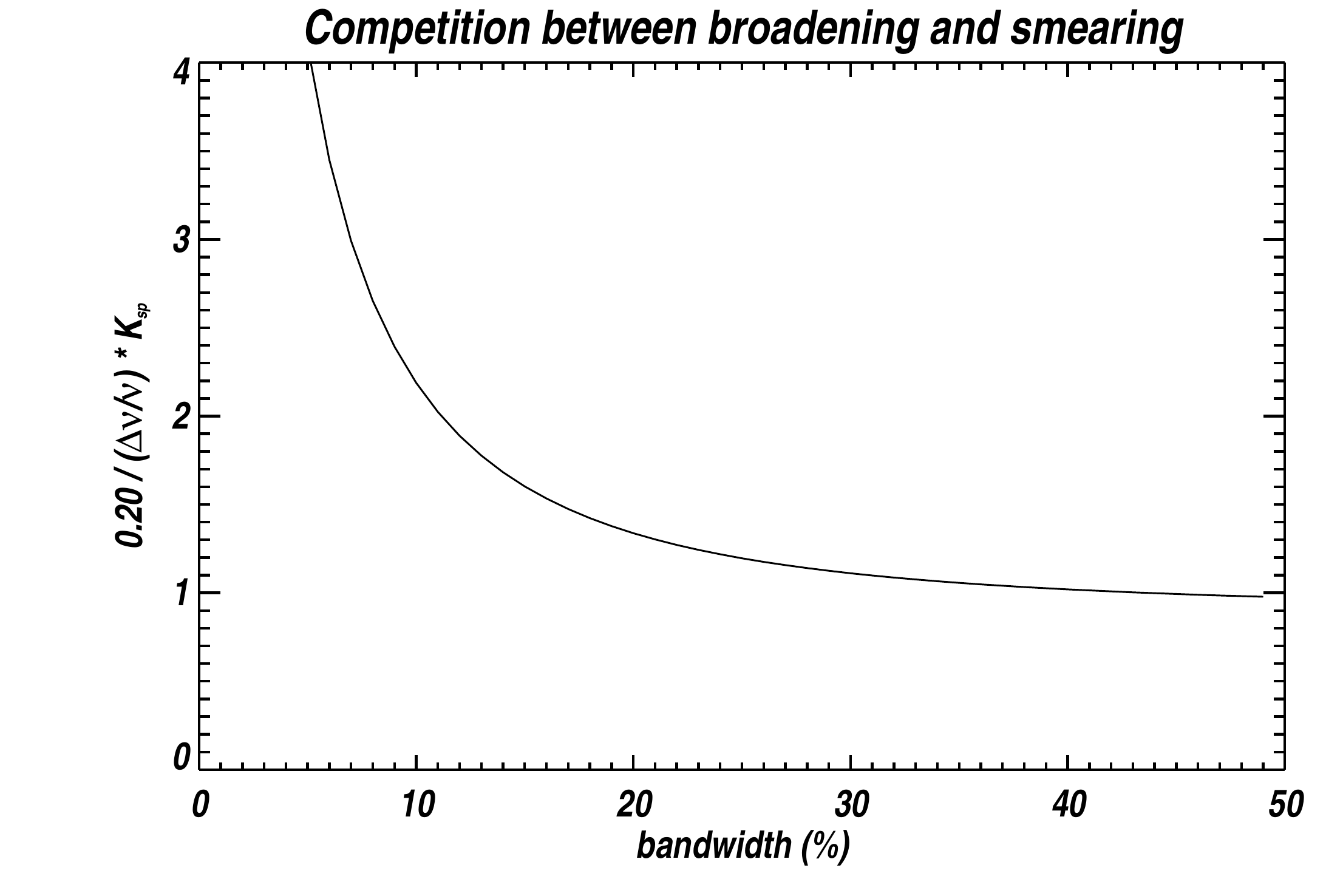}}}
\caption{\small \textit{Left:} sensitivity degradation on $C_l$
extraction due to bandwidth smearing for multipoles between 50 and
200; the quantity $K_{sp}$ is plotted for a dedicated B-mode bolometric
interferometer with a 15-degree FWHM primary beam, for respectively
$\Delta\nu / \nu_0 = $ 10, 20 and 30\% bandwidth.\textit{ Right:} the
quantity $(0.20/ \Delta \nu) \times K_{sp}(l)$, showing the
competition between smearing and broadening, is plotted as a function of
bandwidth for $l=100$, for the same B-mode bolometric interferometer.}
\label{fig:spf} 
\end{figure*}

\section{Loss in sensitivity for CMB experiments \label{sec:sensitivity}}
We have shown in sections~\ref{sec:broadbolvis} and~\ref{sec:virtsub}
how the broadband visibilities defined in section~\ref{sec:broadvis}
can be reconstructed, whatever the frequency dependence of modulation
phase shifters, from the bolometer sequences measured by a broadband
bolometric interferometer. In order to evaluate the resulting loss in
sensitivity for a dedicated CMB experiment, it is important to
understand that it is meaningless to compare \textit{directly}
monochromatic visibilities and broadband ones, because they are not
the same observables (as it is meaningless to directly compare two
signals that have been convolved with kernels of different shape
and/or size). A correct way to deal with this problem is to compare the
sensitivities achieved on the observable of physical interest, here
the CMB power spectra. In this section, we generalize the estimator
introduced in [H08] and derive new formulas for the sensitivity on the
CMB BB power spectrum.

\subsection{Generalization of the Pseudo-Power spectrum estimator}
We make the same assumptions and follow exactly the same arguments
as in [H08], substituting the kernel $\tilde{\beta}(\vec{u},\vec{w})$ for
the Fourier transform of the beam
$\tilde{A}(\vec{w})$.
Assuming perfect E/B separation, one can show that
\begin{equation}
\langle V_B(\vec{u}) V^{\star}_B(\vec{u'})\rangle = \delta(\vec{u}-\vec{u'}) \times \int C_l^{BB}(\vec{w}) \left| \tilde{\beta}(\vec{u}, \vec{w}) \right|^2  d\vec{w}.  \label{eq:vissqmod}
\end{equation}
Recall that in the flat-sky approximation, $l \backsimeq 2 \pi \vert
\vec{u} \vert $ (see e.g.~[\cite{White:1997wq}]). Assuming the power to
be flat enough to be taken out of the integral (actually we rather
assume that $l^2 C_l$ is flat in the simulation, see section
\ref{sec:mcsimu}), in the presence of noise,
\begin{equation}
\langle V_B(\vec{u}) V^{\star}_B(\vec{u'})\rangle = \delta(\vec{u}-\vec{u'}) \ C_l^{BB} \int  \left| \tilde{\beta}(\vec{u}, \vec{w}) \right|^2  d\vec{w} + N(\vec{u},\vec{u'})
\end{equation}
We show in appendix~\ref{app:kernelsquaremod} that the integral of the square modulus of the convolution kernel in the uv-plane actually equals half of the effective beam defined in Eq.~(\ref{eq:effbeam}) :
 \begin{equation}
 \int \left| \tilde{\beta}(\vec{u}, \vec{w}) \right|^2  d\vec{w} = \frac{\Omega_s}{2}=\frac{\Omega}{2\kappa_1}.
\end{equation}
There is a perfect analogy with the monochromatic case where $\frac{\Omega}{2} = \int \left| \tilde{A}(\vec{w}) \right|^2  d\vec{w}$. The simplest unbiased estimator of the power spectrum for a broadband interferometer, in presence of noise, is thus
\begin{equation}
\widehat{C_l} = \frac{2}{\Omega_s} \times \frac{1}{N_{\neq}(l)} \sum_{\beta=0}^{N_{\neq}(l)-1} \left[V(\vec{u_\beta}) V^{\star}(\vec{u_\beta}) - N(\vec{u_\beta},\vec{u_\beta}) \right]. \label{eq:estimatorcl}
\end{equation}

Here $N_{\neq}(l)$ is the number of different modes probing
a given l. It is thus the ratio of the available surface of a
bin $\pi \vec{u} \Delta\vec{u}$ to the effective surface of the kernel in the uv-plane $\Omega_s/2$:
\begin{equation}
N_{\neq}(l)=\frac{2 \pi \vec{u} \Delta\vec{u}}{\Omega_s}= 2l\Delta l \frac{f_{sky}} {\kappa_1}.
\end{equation}

The variance of the estimator for a broadband interferometer can be derived as in [H08], leading to the error on the power spectrum,
\begin{equation}
\Delta C_l^{BI, HI} = \sqrt{\frac{2 \kappa_1}{2l\Delta l f_{sky}}} \left( C_l + \frac{2 \sigma^2_V \kappa_1}{\Omega} \right)
\end{equation}
The only differences with the monochromatic interferometer formula are the $\kappa_1$ factors.

Using the expression for $\sigma_V$ given by Eq.~(\ref{eq:sigmavbi}),
the error on the angular power spectrum measured by a broadband
bolometric interferometer during a time $t$ can finally be written:
\begin{equation}
\Delta C_l^{BI} = \sqrt{\frac{2 \kappa_1}{2l\Delta l f_{sky}}} \left( C_l + \frac{2  N_h \mathrm{NET}^2_{BI} \Omega}{N_{eq}^2 \ t} \kappa_1 \right). \label{eq:sensfirst}
\end{equation}
The $\mathrm{NET_{BI}}$ of bolometers scales as the inverse square root of the bandwidth, $\mathrm{NET_{BI}} \propto 1/\sqrt{\Delta \nu}$ (cf. appendix~\ref{app:net}); we make it appear explicitly by writing
\begin{equation}
 \mathrm{NET}^2_{BI}  =  \mathrm{NET}^2_{BI, 20\%} \frac{0.20}{(\Delta \nu/\nu_0)} ,
\end{equation}
where $\mathrm{NET}^2_{BI, 20\%}$ is the Noise Equivalent Temperature of 20\%-bandwidth bolometers. Eq.~(\ref{eq:sensfirst}) becomes:
\begin{equation}
\Delta C_l^{BI} = \sqrt{\frac{2}{2l\Delta l f_{sky}}} \left( \sqrt{\kappa_1} C_l + \frac{2  N_h \mathrm{NET}^2_{BI, 20\%}\  \Omega}{N_{eq}^2 \ t} \times \frac{0.20}{(\Delta \nu/\nu_0)} K_{sp}(l) \right),
\label{eq:sensfinal}
\end{equation}
where the \textit{smearing penalty factor} is defined by:
\begin{equation}
K_{sp}(l) = \kappa_1^{3/2} =  \left( 1 + \frac{\Omega \left(\sigma_\nu / \nu_0 \right)^2 l^2 }{2\pi }  \right)^{3/4}.
\end{equation}
If we neglect the $\sqrt{\kappa_1}$ penalty on the sample variance,
the factor of sensitivity degradation due to bandwidth smearing for a
bolometric interferometer is indeed given by $K_{sp}(l)$. The physical
interpretation of Eq.~(\ref{eq:sensfinal}) is straightforward: the
sensitivity improvement due to bandwidth broadening (more photons are
collected) is in competition with the sensitivity degradation due to
bandwidth smearing (the fringes are degraded). Figure~\ref{fig:spf}
(right) shows the evolution of $(0.20/ \Delta \nu) \times K_{sp}(l)$
as a function of bandwidth. We see that, for the typical B-mode
experiment considered, the smearing begins to cancel the broadening
for bandwidths larger than $20\%$ -which is fortunately the typical
bandwidth of bolometers used in CMB experiments\footnote{For
ground-based experiments, atmospheric emission lines exclude the
possibility of larger bandwidth anyway.}. Figure~\ref{fig:spf} (left)
shows that the total loss in sensitivity on power spectra due to
bandwidth smearing is about $2$ for $l=150$, for a typical dedicated
B-mode experiment with 20\% bandwidth. This result, which may seem
unexpected considering the bad reputation of radio-interferometers
concerning bandwidth, is mainly due to the fact that the spatial
resolution required for the observation of CMB angular correlations is
poorer than that required for the observation of point sources.

\subsection{Inefficiency of the reconstruction in sub-bands for beating bandwidth smearing}
If the visibilities were reconstructed in $n_{vsb}$ sub-bands, the smearing would be reduced but the signal to noise ratio in each $l$ band would be decreased, leading to a $\sqrt{n_{vsb}}$ additionnal factor in $\sigma_V$ as explained in section \ref{sec:virtsub}. The smearing penalty factor would be
\begin{equation}
K^{SB}_{sp}(l) = {n_{vsb}} \ \kappa_1^{3/2} = n_{vsb} \ \left( 1 + \frac{\Omega \left(\sigma_\nu / \nu_0 \right)^2 l^2 }{2\pi \ n_{vsb}^2 }  \right)^{3/4}.
\end{equation}
However, it can be checked that $K^{SB}_{sp}(l)$ is always greater than $K_{sp}(l)$, whatever the number of sub-bands, meaning that the loss in signal to noise ratio is always more penalizing than the smearing reduction. 

\begin{figure*}[!ht]
\centering\resizebox{\hsize}{!}{\centering{
\includegraphics[angle=0]{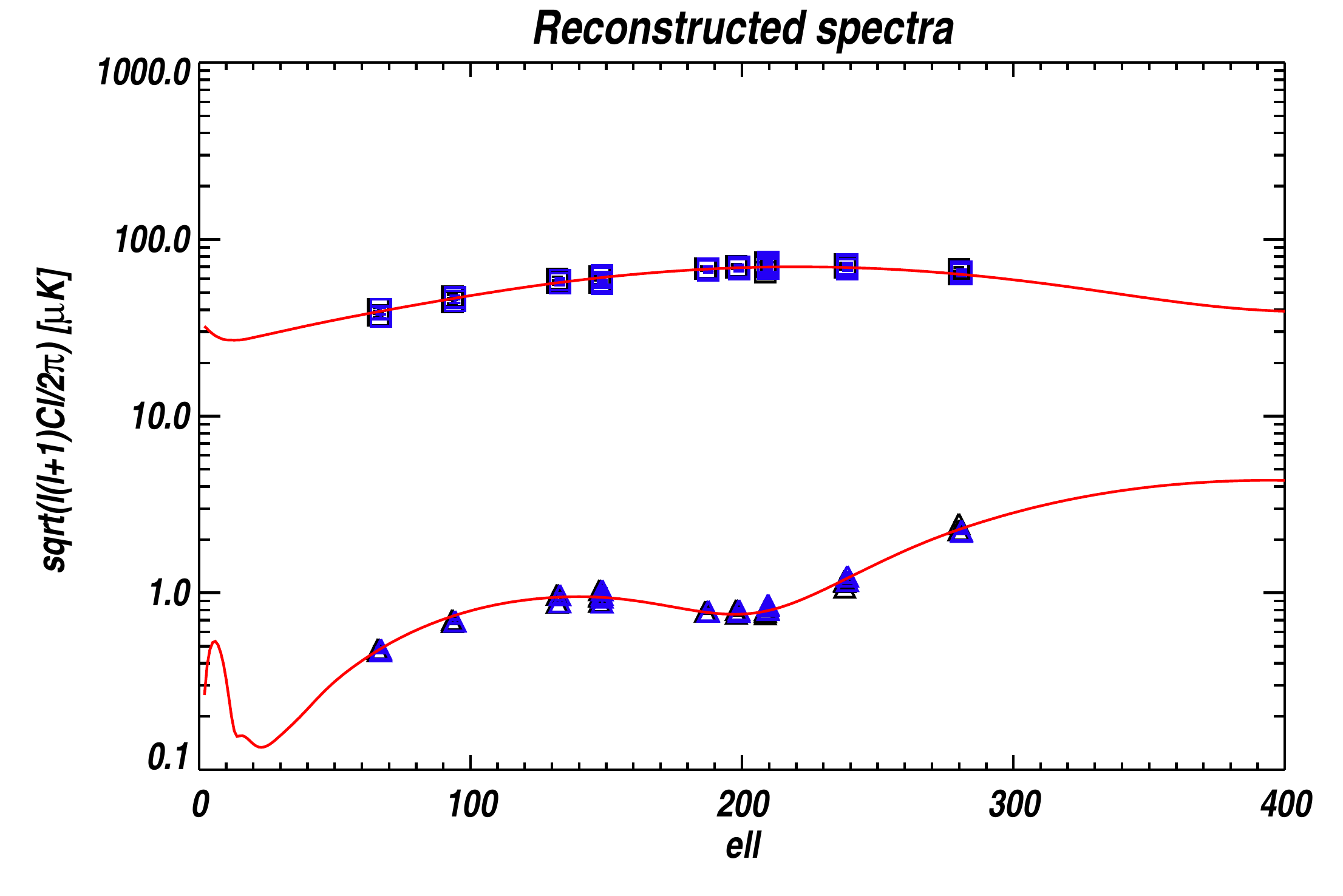} \includegraphics[angle=0]{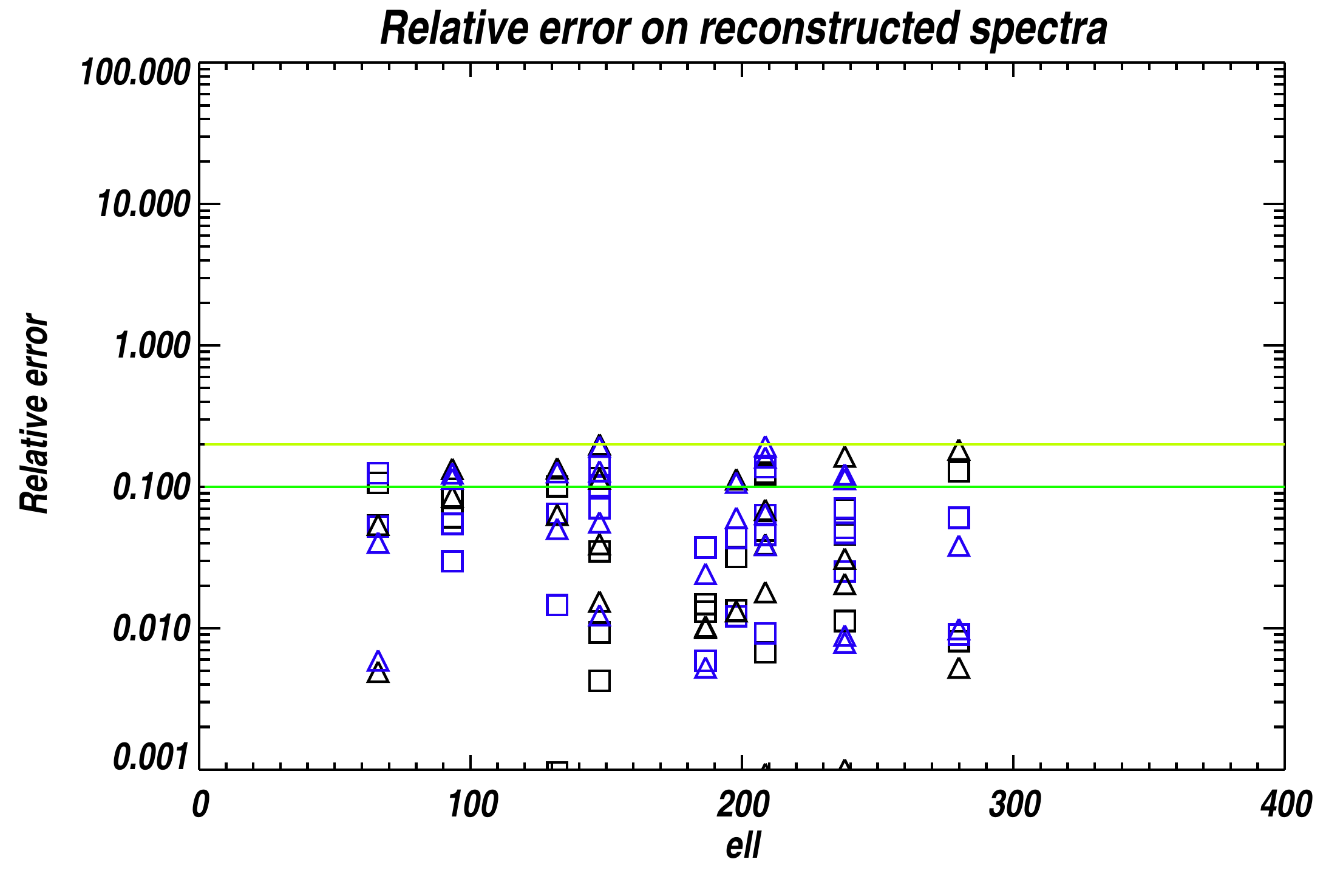}}}
\caption{\small \textit{Left:} Reconstructed temperature and polarization power spectra for a monochromatic (black squares) and a $20\%$ broadband (blue squares) bolometric interferometer (16 horns with 15 degrees FWHM primary beam, phase shifters constant with respect to frequency, 12 modulation phase shift values, no instrumental noise and $N_{maps}=100$ in each case). Red lines show the input theoretical spectra. \textit{Right:} Relative error on the reconstruction $(\hat{C}^{MC}_l-C_l)/C_l$ (squares for temperature spectra, triangles for polarization ones, black for monochromatic, blue for broadband). The green and yellow lines shows the expected error levels $1 / \sqrt{N_{maps}}$ at one and two sigmas.}
\label{fig:recspec_fps} 
\end{figure*}
 
\subsection{Comparison with an imager and a heterodyne interferometer}
We now correct the ratio formulas derived in [H08]. The comparison
between the sensitivities of a heterodyne and a bolometric
interferometer is not straigthforward since there is an important
difference of hardware design between the two kind of interferometry
regarding bandwidth. In a radio heterodyne interferometer such as
DASI~[\cite{DASI}] or CBI~[\cite{CBI}], the analog correlators only
work at low frequencies (typically below 2 GHz), so the broadband
signal collected by each horn is divided into different channels of
typically 1 GHz bandwidth each and then downconverted before being
correlated. This forced division has prevented past interferometer CMB
experiments from any important bandwidth smearing effects, but the
price to pay was of course the hardware complexity of such
systems. Bolometers are in the other hand naturally broadband, and we
have shown in this article how the monochromatic bolometric
interferometer described in [C08] generalizes almost naturally into a
broadband bolometric interferometer: a broadband instrument only needs
broadband components (horns, filters, etc.). To correct the ratio
formula obtained in [H08], one can neglect the bandwidth smearing for
heterodyne interferometers (for a $1\%$ bandwidth, $\kappa_1$ is
always very close to 1); the formula is then only corrected by the
smearing penalty factor:
\begin{equation}
\frac{\Delta C^{HI}_l}{\Delta C^{BI}_l} =\left( \frac{N_{eq}}{N_h} \right) \times \frac{\mathrm{NET_{HI}}}{\mathrm{NET_{BI}}} \times K_{sp}^{-1}.
\end{equation}
To be completely fair, one must keep in mind that in the actual state of the technological art, it seems difficult to design heterodyne interferometers with bandwidth larger than $10\%$.

The comparison between a bolometric interferometer and an imager is also only modified by the smearing penalty factor. If the experiment is dominated by instrumental noise, the ratio of the variances becomes
\begin{equation}
\frac{\Delta C^{Im}_l}{\Delta C^{BI}_l} =\left( \frac{N_{eq}}{N_h} \right)^2 \times \frac{1}{B^2_l} \times K_{sp}^{-1}.
\end{equation}

\section{Monte-Carlo simulations \label{sec:mcsimu}} 
We have performed a Monte-Carlo simulation in order to check the
results obtained in this article. The basic principle is, starting
from CMB maps generated from theoretical spectra, to compute the
sequences of data measured by a broadband bolometric interferometer,
to reconstruct visibilities from these sequences, and to estimate
power spectra from these visibilities. The comparison between input
and output spectra then allows us to check the analytical calculations
(and the associated assumptions) of this article. The code is
available upon request; questions or comments can be adressed by
e-mail to the authors.

\subsection{Simulation overview}

We consider a ``standard" bolometric interferometer (as defined in
section \ref{sec:broadbolvis}) constituted by a square array of $N_h$
horns, the associated $N_h$ polarization splitters, the $2N_h$
modulation phase shifters, a beam combiner and $N_{out}$
bolometers. In this simulation we only compute the power measured by
one of the bolometers. We assume that the beams of the horns are all
described by the same perfect gaussian function of
Eq.~(\ref{eq:gaussianbeam}). We assume that the phase shift values
taken by the modulation phase shifters are equally spaced.
The physical input parameters are then the number of horns $N_h$, the
horns' radius, the distance between two adjacent horns, the FWHM of the
gaussian beam, the number of phase shift values taken by the
modulation phase shifters $N_{\phi}$, the central observation
frequency $\nu_0$, the bandwidth $\Delta \nu$, the form of the
bandpass function (either gaussian or top hat) $J$, and the number of
data samples in one sequence $N_d$.

The CMB maps are generated from spectra given by the standard WMAP-5
cosmological model (although the spectra's shapes are not really
important for this simulation, the only crucial feature being the
amplitude ratio between temperature and polarization spectra). The
question of the E and B modes separation in interferometry is beyond
the scope of this article and of this simulation. So we only consider
the TT and EE spectra, which we call from now respectively the
temperature and polarization spectra. Computation of the $\beta$ and
$\eta$ kernels involves a numerical integration over the frequency
band, while computation of the samples measured by the bolometer
involves one over the uv-plane. The numerical input parameters are
then the resolution in the uv-plane, the resolution in the frequency
band, and the number of CMB map realisations $N_{maps}$.

The simulation pipeline is the following:
\begin{enumerate}
\item The position of primary horns and the associated set of baselines are generated. 
\item The $\beta$ and $\eta$ kernels are computed -- either from
analytical formulas (given by Eq.~(\ref{eq:fourthbeta}) and
Eq.~(\ref{eq:etakernana}) or numerical integrations (following
Eq.~(\ref{eq:convkernap}) and Eq.~(\ref{eq:firsteta})), which enables us
to check the analytical formulas -- for every different baseline and
for every phase difference in the case of $\eta$.
\item Begining of MAPS loop. CMB temperature and polarization maps are generated from theoretical spectra. 
\item Monochromatic and broadband visibilities are computed by
convolving the maps' Fourier transforms and the $\beta$ kernels for
every baseline. ``Generalized" broadband visibilities
(i.e. the convolution of maps' fourier transforms and $\eta$ kernels)
are computed for every baseline and for every phase
difference.
\item Random phase sequences are generated for every horn and both
polarizations, respecting the coherent summation of equivalent
baselines scheme described in [C08]. Phase differences are then
computed for every baseline.
\item A sequence of $N_d$ data sets $S_k$ measured by the bolometer is computed (Eq.~(\ref{eq:bolosignal1})) by summing ``generalized" broadband visibilities, following the phase sequences.
\item The modulation matrix is generated (see [C08] for its explicit expression in the monochromatic case). 
\item The visibilities are reconstructed by solving the linear problem of Eq.~(\ref{eq:linprobfps1}). End of MAPS loop.
\item Measured spectra are computed from the reconstructed visibilities using the estimators defined in section \ref{sec:sensitivity}. Relative errors are obtained by comparison with the input theoretical spectra. 
\end{enumerate}

\subsection{Validation of the work principle of a broadband bolometric interferometer and test of the broadband estimator}
In step 9, we actually do not average the modulus of the reconstructed
visibilities over the different baselines matching the same multipole
as in Eq.~(\ref{eq:estimatorcl}), but for each different baseline, we
average over all the $N_{maps}$ maps realisations. Moreover, the power
is actually not flat over the $\Delta l$ of integration, and cannot be
taken out of the integral in Eq.~(\ref{eq:vissqmod}), but we can use
the fact that $l^2 C_l$ is nearly flat to define an unbiased
estimator (assuming no instrumental noise) for the broadband
interferometer:
\begin{equation}
\widehat{C_l}^{MC} = \frac{2}{\Omega_s^{MC}} \times \frac{1}{N_{maps}} \sum_{m=0}^{N_{maps}-1} V_m(\vec{u_{\beta}}) V_m^{\star}(\vec{u_{\beta}}).
\end{equation}
where $V_m(\vec{u_{\beta}})$ is the broadband visibility defined for the baseline $\vec{u_{\beta}}$ ($l=2 \pi \vert \vec{u_{\beta}} \vert$) and the map $m$ and where
\begin{equation}
\Omega_s^{MC} = \int \frac{\vert \vec{w} \vert^2}{\vert \vec{u_{\beta}} \vert^2} \vert \tilde{\beta}(\vec{u_{\beta}}, \vec{w})\vert^2 d\vec{w} \sim \frac{\Omega}{2\kappa_1}.
\end{equation}

We thus expect the spectra reconstruction to be only affected by the ``sample" variance: 
\begin{equation}
\Delta C_l^{MC} = C_l / \sqrt{N_{maps}} .
\end{equation}

\begin{figure*}[!ht]
\centering\resizebox{\hsize}{!}{\centering{
\includegraphics[angle=0]{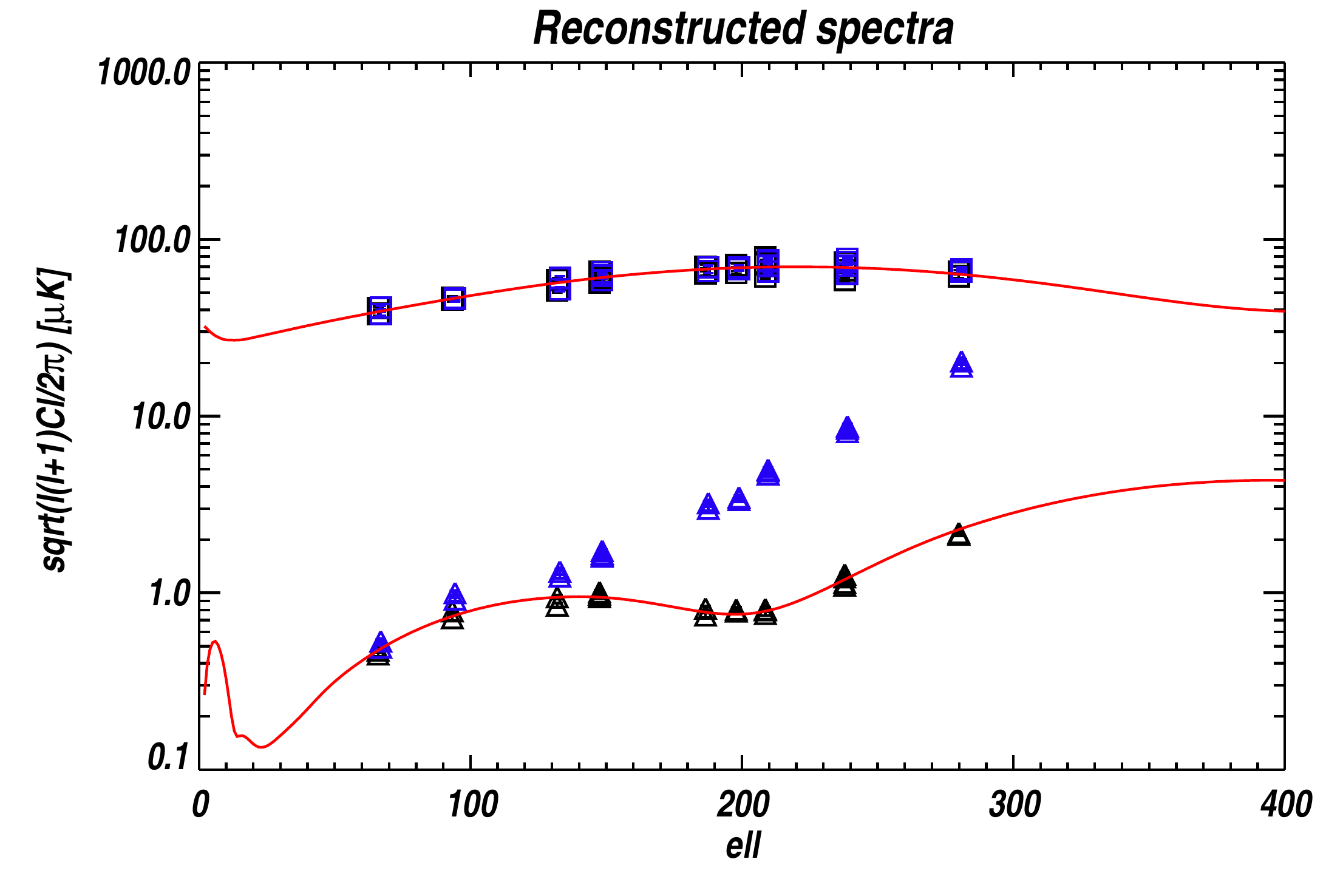} \includegraphics[angle=0]{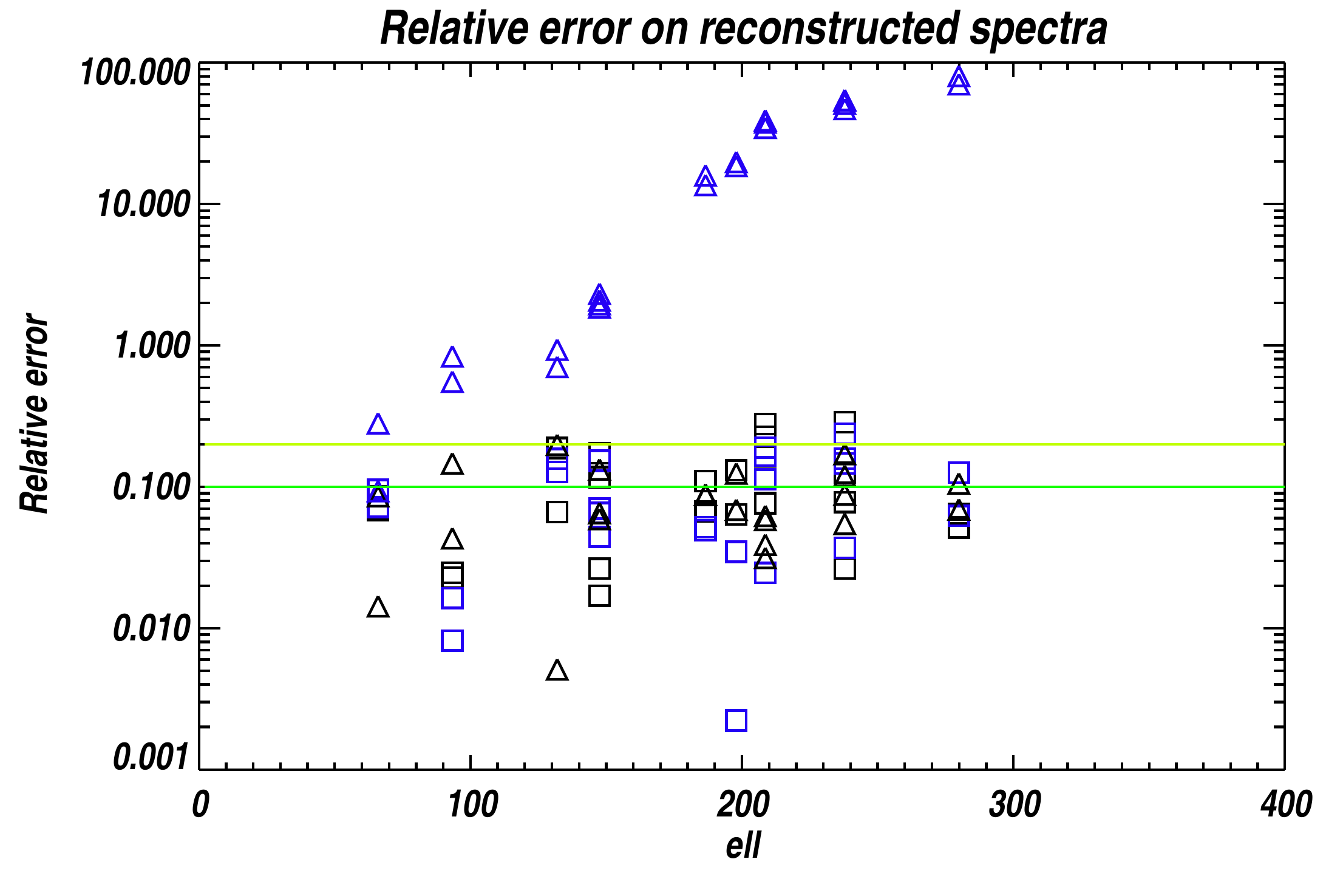}}}
\caption{\small \textit{Left:} Phase shifters linear with respect to frequency and reconstruction with monochromatic modulation matrix. Reconstructed temperature and polarization power spectra for a monochromatic (black squares and triangles) and a $20\%$ broadband (blue squares and triangles) bolometric interferometer (16 horns with 15 degrees FWHM primary beam, 12 modulation phase shift values, no instrumental noise and $N_{maps}=100$ in each case). Red lines show the input theoretical spectra. \textit{Right:} Relative error on the reconstruction $(\hat{C}^{MC}_l-C_l)/C_l$ (squares for temperature spectra, triangles for polarization ones, black for monochromatic, blue for broadband). The green and yellow lines shows the expected error levels $1 / \sqrt{N_{maps}}$ at one and two sigmas.}
\label{fig:recspec_ldps} 
\end{figure*}

\begin{figure*}[!ht]
\centering\resizebox{\hsize}{!}{\centering{
\includegraphics[angle=0]{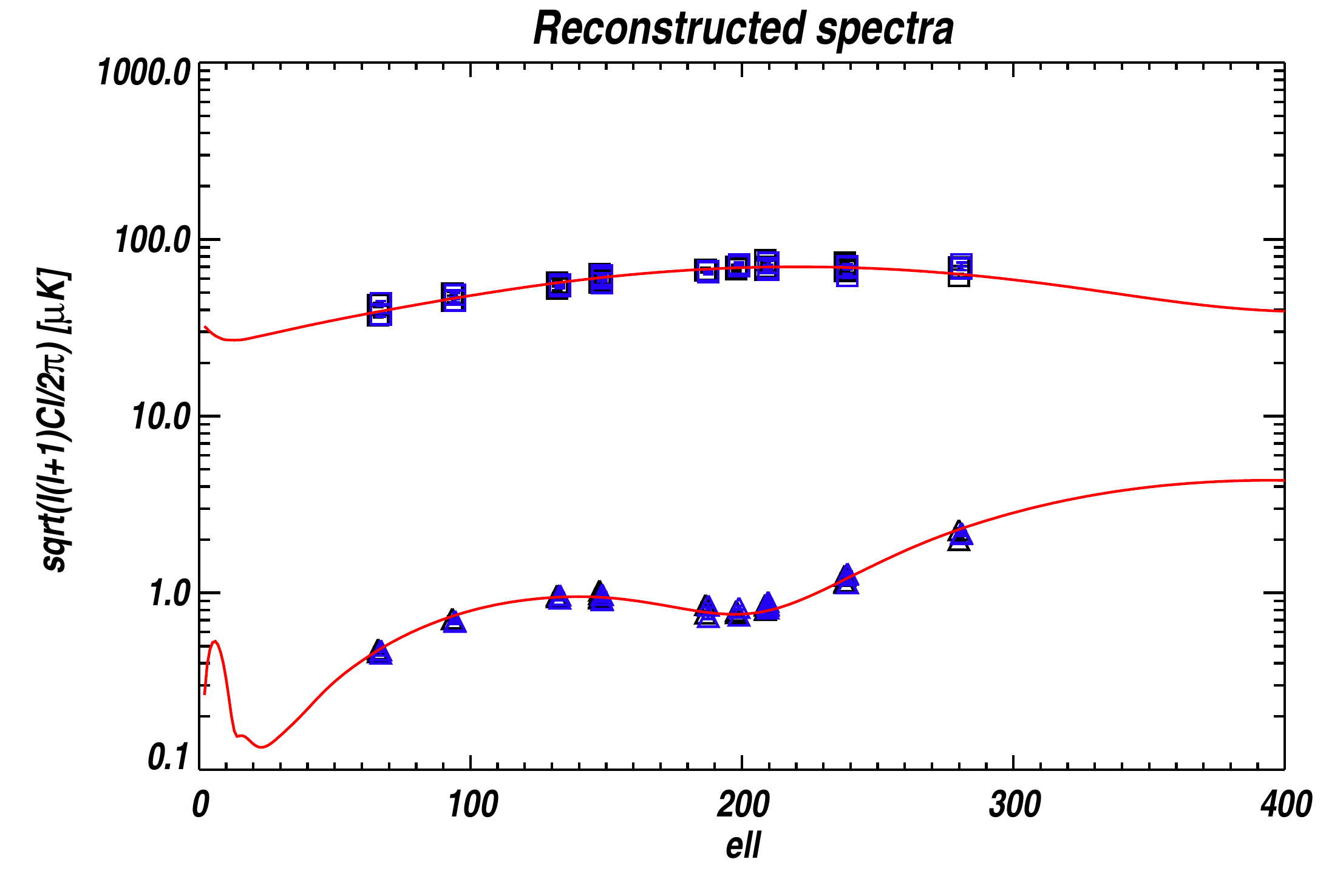} \includegraphics[angle=0]{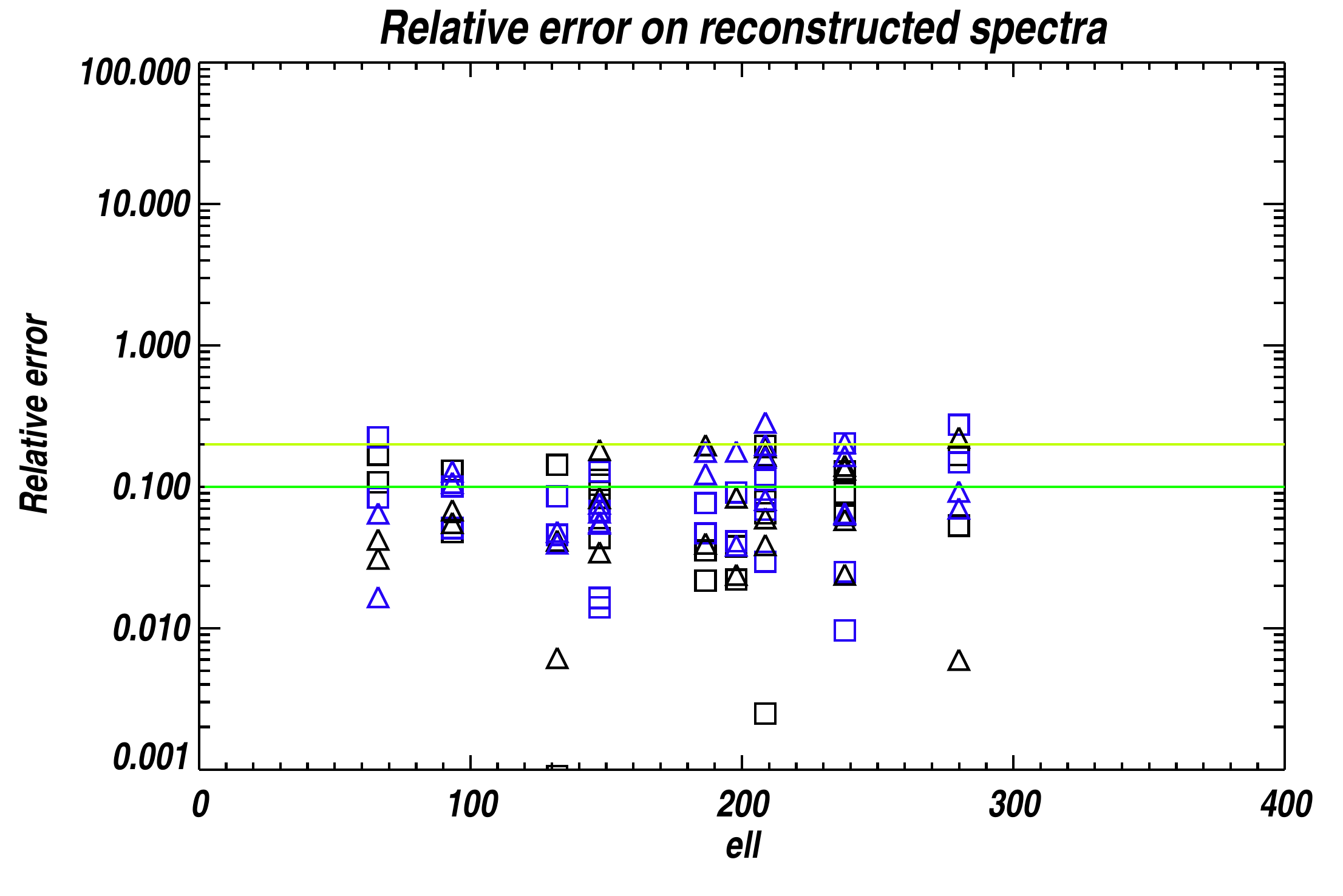}}}
\caption{\small \textit{Left:} Phase shifters linear with respect to frequency and reconstruction with the extended modulation matrix ($5$ virtual sub-bands for $I$ visibilities only). Reconstructed temperature and polarization power spectra for a monochromatic (black squares and triangles) and a $20\%$ broadband (blue squares and triangles) bolometric interferometer (16 horns with 15 degrees FWHM primary beam, 15 modulation phase shift values, no instrumental noise and $N_{maps}=100$ in each case). Red lines show the input theoretical spectra. \textit{Right:} Relative error on the reconstruction $(\hat{C}^{MC}_l-C_l)/C_l$ (squares for temperature spectra, triangles for polarization ones, black for monochromatic, blue for broadband). The green and yellow lines shows the expected error levels $1 / \sqrt{N_{maps}}$ at one and two sigmas.}
\label{fig:recspec_ldps_RVSB} 
\end{figure*}

We run the simulations with the following input parameters: 16 horns,
15 degrees FWHM primary beam, 
12 modulation phase shift values, a 90 GHz central
frequency, 20\% bandwidth, $N_{maps}=100$. We first simulate the case
of phase shifters constant with respect to frequency (the
$\eta$-kernels are computed following Eq.~(\ref{eq:firsteta}) and
Eq.~(\ref{eq:etakernana})) and use the monochromatic modulation matrix
defined in [C08] to reconstruct the visibilities. The results, shown
in figure~\ref{fig:recspec_fps}, validate our study since our
broadband estimator, taking the smearing into account, 
reconstructs the generated power spectra well: a broadband bolometric
interferometer with phase shifters constant with respect to frequency
works exactly like a monochromatic one, but the reconstructed
visibilities are the predicted smeared ones. We then simulate the case
of frequency-dependent modulation phase shifters, with kernels 
computed following Eq.~(\ref{eq:firsteta}) and
Eq.~(\ref{eq:secondeta}). Figure~\ref{fig:recspec_ldps} shows the
dramatic effect described in subsection~\ref{subsec:fdps} on the
reconstructed polarization spectrum when the monochromatic modulation
matrix is used to reconstruct the
visibilities. Figure~\ref{fig:recspec_ldps_RVSB} shows the efficiency
of using the extended modulation matrix described in
subsection~\ref{subsec:spescheme} to reconstruct the polarization
visibilities: the intensity visibilities have been reconstructed in 5
sub-bands, completely removing the error on the polarization
visibilities reconstruction in the configuration considered (at the
level of sample variance considered of course).

\section{Conclusion}
In this article we have analytically and numerically studied the work
principle of a broadband bolometric interferometer. We have defined
its (indirect) observables -- the broadband visibilities -- and introduced
numerical methods to reconstruct them. We have finally calculated the
sensitivity of such an instrument dedicated to the B-mode.

Bolometers are naturally broadband, and consequently the design of a
broadband bolometric interferometer is identical to the design of the
monochromatic one described in [C08]: a broadband bolometric
interferometer only needs broadband components (horns, filters,
etc.). Nevertheless, we have seen that the modulation matrix that
should be used to reconstruct the broadband visibilities depends on
some hardware properties of the modulation phase shifters. If these
are constant with respect to frequency, the modulation matrix should
be the one defined in [C08] for a monochromatic instrument. If they
are dependent on frequency (this dependence should be
known of course), a more complicated scheme involving a reconstruction
in sub-bands of the intensity visibilities should be
performed. We have checked with a numerical simulation that in
both cases the visibilities can be reconstructed without other loss
in sensitivity than the one due to the smearing.

Visibilities are defined as the convolution of the Fourier transform
of the signal with a kernel, which is defined as the Fourier transform
of the primary beam in the monochromatic case. We have shown that the
effect of the smearing is to stretch this kernel, in the baseline
direction only, and that the amplitude of the smearing only depends on
three quantities: the bandwidth, the baseline length and the size of
the primary beam. We have finally defined, as a function of broadband
visibilities, a new power spectrum estimator and derived from it a
generalized uncertainty formula.

The main conclusion of this article is that for a bolometric interferometer dedicated to CMB B-mode, the sensitivity loss -due to bandwidth smearing- is quite acceptable (a factor 2 loss for a typical 20\% bandwidth experiment).

\begin{acknowledgements}
      The authors are grateful to the whole QUBIC collaboration for fruitful discussions.
\end{acknowledgements}

\begin{appendix}
\section{Analytical derivation of the $\tilde{\beta}$-kernel \label{app:betakernel}}
The kernel of Eq.~(\ref{eq:firstbeta}) can be written as a gaussian integral:
\begin{eqnarray} 
\tilde{\beta}_{ap}(\vec{u_0}, \vec{w}) &=&  \frac{\Omega}{\sigma_\nu \sqrt{2\pi}} e^{\frac{B^2}{4A}-C} \int e^{-A(\nu-\frac{B}{2A})^2 } d\nu \\
 &=&  \frac{\Omega}{\sigma_\nu \sqrt{2\pi}} \exp \left[ \frac{B^2}{4A}-C\right] \sqrt{\frac{\pi}{A}} ,
\end{eqnarray}
where we have defined the quantities
\begin{eqnarray}
A &=& \frac{1}{2\sigma_\nu^2} + \pi \Omega \frac{\vec{u_0}^2}{\nu_0^2}, \\
B &=& \frac{\nu_0}{\sigma_\nu^2} + 2 \pi \Omega  \frac{\vec{u_0} \cdot (\vec{u_0} - \vec{w'})}{\nu_0}, \\
C &=& \frac{\nu_0^2}{2\sigma_\nu^2} + \pi \Omega (\vec{u_0}- \vec{w'})^2,
\end{eqnarray}
and where we have made the variable substitution
\begin{equation}
\vec{w'} = \vec{u_0}-\vec{w}.
\end{equation}
It is straightforward to show that
\begin{equation}
 \frac{\Omega}{\sigma_\nu \sqrt{2\pi}} \sqrt{\frac{\pi}{A}} = \frac{\Omega}{\sqrt{1+2\pi\Omega \left(\frac{\sigma_\nu}{\nu_0} \right)^2 \vec{u_0}^2 }},
\end{equation}
\begin{equation}
\frac{B^2}{4A}-C = -\pi \Omega \vec{w'}^2 + \frac{2\pi^2 \Omega^2 \left( \frac{\sigma_\nu}{\nu_0}\right)^2} {1+2\pi \Omega \left( \frac{\sigma_\nu}{\nu_0}\right)^2 \vec{u_0}^2}(\vec{u_0} \cdot \vec{w'})^2.
\end{equation}
Thus, we can write the kernel in function of the Fourier transform of the beam as in Eq.~(\ref{eq:secondbeta}).

\section{Analytical derivation of the $\tilde{\eta}$-kernel \label{app:etakernel}}
The kernel of Eq.~(\ref{eq:firsteta}) can be written as the Fourier transform of a gaussian:
\begin{eqnarray}
  \tilde{\eta}^ {BI}_{b,p}(\vec{u_0}, \vec{w}) &=& \Omega e^{-\frac{B^2}{4A}+C} \int e^{-A(\nu-\frac{B}{2A})^2} \ e^{i \Delta \Phi_{b,p} \frac{\nu}{\nu_0} } d\nu \\
   &=&  \Omega e^{-\frac{B^2}{4A}+C} \sqrt{\frac{\pi}{A}} \ e^{-\Delta \Phi_{b,p}^2 \frac{1}{4 \nu_0^2 A} } \ e^{i \Delta \Phi_{b,p} \frac{B}{2\nu_0 A}}   ,
\end{eqnarray}
where A, B and C are the quantities defined in appendix A. It is straightforward to show that
\begin{eqnarray}
G & = & \frac{1}{4 \nu_0^2 A} = \left(\frac{\sigma_\nu}{\nu_0}\right)^2 \times \frac{1}{2\kappa_1^2}, \\
\frac{B}{2\nu_0 A} &=& 1 - \frac{2\pi\Omega \vec{u}_0 \cdot \vec{w'} (\frac{\sigma_\nu}{\nu_0})^2}{1 + 2\pi \Omega \vec{u_0}^2 (\frac{\sigma_\nu}{\nu_0})^2} = 1-H(\vec{w'}),
\end{eqnarray}
with
\begin{eqnarray}
H(\vec{w'}) &=& \left(1 - \frac{1}{\kappa_1^2}\right) \times \frac{\vec{u_0} \cdot \vec{w'}}{\vec{u_0}^2}.
\end{eqnarray}
And we can finally write the kernel as in Eq.~(\ref{eq:secondeta}).

\section{Integration of the $\tilde{\beta}$-kernel square modulus in uv-plane \label{app:kernelsquaremod}}
We calculate this integral using the approximate kernel of Eq.~(\ref{eq:thirdbeta}). The variable substitution $\vec{w'} = \vec{u_0}-\vec{w}$ does not change the integral:
\begin{equation}
\int \left\vert \tilde{\beta}_{ap}(\vec{u}, \vec{w}) \right\vert^2 d\vec{w} = \int \left\vert \tilde{\beta}_{ap}(\vec{u}, \vec{w'}) \right\vert^2 d\vec{w'}.
\end{equation}
Using Parseval's theorem, one gets
\begin{equation}
 \int \left\vert \tilde{\beta}_{ap}(\vec{u}, \vec{w'}) \right\vert^2 d\vec{w'} =  \int \left\vert \beta_{ap}(\vec{u}, \vec{n}) \right\vert^2 d\vec{n},
\end{equation}
where $\beta_{ap}$ is the inverse Fourier transform of  $\tilde{\beta}_{ap}$. Because $\tilde{\beta}_{ap}(\vec{u}, \vec{w'})$ is a positive gaussian function of $ \vec{w'}$,
\begin{equation}
\int \left\vert \beta_{ap}(\vec{u}, \vec{n}) \right\vert^2 d\vec{n} = \frac{1}{2} \int \left\vert \beta_{ap}(\vec{u}, \vec{n}) \right\vert d\vec{n} = \frac{1}{2}\tilde{\beta}_{ap}(\vec{u}, \vec{w'}=\vec{0}).
\end{equation}
By definition $\Omega_s = \tilde{\beta}_{ap}(\vec{u}, \vec{w'}=\vec{0})$, so finally
\begin{equation}
\int \left\vert \tilde{\beta}_{ap}(\vec{u}, \vec{w}) \right\vert^2 d\vec{w} = \frac{\Omega_s}{2}.
\end{equation}

\section{Noise Equivalent Power and Noise Equivalent Temperature \label{app:net}}
The spectral power $I_\nu$ collected by a horn of surface $S$ is defined in Eq.~(\ref{eq:specpower}). We assume for simplicity that the number of bolometers equals the number of horns. The total power measured by a bolometer is then:
\begin{eqnarray}
P_{tot} &=& \iint I_\nu(\vec{n}) J(\nu-\nu_0) A_\nu(\vec{n}) d\vec{n} d\nu\\
& \simeq & S \Omega \ J(0) \ \Delta \nu \int \frac{\nu_0^2}{\nu^2}B_\nu \ J_N(\nu-\nu_0) d\nu.
\end{eqnarray}

The Noise Equivalent Power (NEP) due to photon noise on a bolometer, in units of [$\mathrm{W.Hz^{-1/2}}$], is given by [\cite{lamarre}]:
\begin{equation}
\begin{split}
\mathrm{NEP^2} &= 2 J(0) \ \Delta \nu \Omega \int h\nu  \frac{\nu_0^2}{\nu^2} I_\nu J_N(\nu-\nu_0) d \nu  \ \ + \ \ ... \\
&  \hspace{2cm}  2 J(0) \ \Delta \nu \Omega \int  \frac{c^2}{2 \nu^2}  \frac{\nu_0^2}{\nu^2} I_\nu^2 J_N(\nu-\nu_0) d \nu .
\end{split}
\end{equation}

For CMB work, bolometer sensitivity is usually quoted as a Noise Equivalent Temperature in units of [$\mathrm{K.s^{1/2}}$], firstly to simplify the comparison with the sensitivity of coherent receivers and secondly to simplify the calculation of sensitivity on $C_l$ since they are defined in temperature units.
The conversion is given by:
\begin{equation}
\mathrm{NET} = \frac{\mathrm{NEP}}{\sqrt{2} (\partial P_{tot}/\partial T)} \label{eq:netdef}
\end{equation}
It is straightforward to show that
\begin{equation}
\frac{\partial P_{tot}}{\partial T} = S \Omega \ J(0) \ \Delta \nu \int  \frac{\nu_0^2}{\nu^2} \frac{\partial B_\nu}{\partial T} \ J_N(\nu-\nu_0) d\nu.
\end{equation}
The NET thus scales as the inverse square root of the bandwidth:
\begin{equation}
\mathrm{NET} \propto \frac{1}{\sqrt{\Delta \nu}}.
\end{equation}

\section{Noise in visiblity measurement in Kelvin \label{app:noisek}}
We can write the following relation between $\sigma_0$ in units of [$\rm W.s^{1/2}$] and the NEP in units of [$\rm W.Hz^{-1/2}$]:
\begin{equation}
\frac{\sigma_0}{\sqrt{N_t}} = \frac{NEP}{\sqrt{2} \sqrt{t_S}},
\end{equation}
where $t_S$ is the duration of one phase sequence. Starting from Eq.~(\ref{eq:noisesigma0}), the noise in Watt on a broadband visibility measured during a time $t$ by an experiment is then
\begin{equation}
\sigma_V [\mathrm{in \ W}] = \alpha \ \frac{\mathrm{NEP}}{\sqrt{2} \sqrt{t} \ J(0) \ \Delta \nu}.
\end{equation}
The quantity $\alpha$ is different depending on whether one is considering an heterodyne or a bolometric interferometer. In the latter case, it has been derived in [C08]: $\alpha=\frac{\sqrt{N_h}}{N_{eq}}$. We have defined the visibilities in temperature units in Eq.~(\ref{eq:tempvisdef}). The noise on a visibility measurement in Kelvin is thus given by
\begin{equation}
\sigma_V [\mathrm{in \ K}] = \sigma_V [\mathrm{in \ W}] \times \frac{1}{ S \left. ({\partial B_\nu}/{\partial T}) \right|_{\nu_0}}.
\end{equation}
Using the definition of the NET given in~\ref{eq:netdef}, the noise on a visibility measurement in Kelvin finally becomes
\begin{equation}
\sigma_V [\mathrm{in \ K}] = \alpha \ \frac{\mathrm{NET} \ \Omega}{\sqrt{t}} \ \kappa_2,
\end{equation}
where we have introduced
\begin{equation}
\kappa_2= \int  \frac{\nu_0^2}{\nu^2} \left( 
\frac{{\partial B_\nu}/{\partial T}}{  \left. ({\partial B_\nu}/{\partial T}) \right|_{\nu_0}} \right) J_N(\nu-\nu_0) d\nu.
\end{equation}
We see in table~\ref{tab:k2} that $\kappa_2 \simeq 1$ is a good approximation.
\begin{table}[!h]
\center
\begin{tabular}{|c|c|c|}
\hline  30 GHz& 90 GHz & 250 GHz\\ 
\hline  $\sim 1-10^{-3}$& $\sim 1-10^{-2}$ & $\sim 1+10^{-2}$ \\ 
\hline 
\end{tabular}
\caption{Values of $\kappa_2$ for a $20\%$ bandwidth, for different central frequencies $\nu_0$. We assume the instrument is observing, through the gaussian bandpass function defined in Eq.~(\ref{eq:gaussbp}), a 3K black body source whose intensity is given by Eq.~(\ref{eq:bbintensity}). }  
\label{tab:k2}
\end{table}

\end{appendix}

\end{document}